\documentclass[a4paper,11pt]{article}
\usepackage{pdfpages}
\usepackage{subcaption}
\usepackage{ifpdf}
\usepackage[toc,page]{appendix}
\usepackage{amsmath}

\usepackage{jheppub} 

\usepackage[T1]{fontenc} 
\usepackage{graphics}
\usepackage{graphicx}
\usepackage{dcolumn}
\usepackage{bm}
\usepackage{mathrsfs}
\usepackage{pstricks}
\usepackage{color}
\usepackage{natbib}
\usepackage[normalem]{ulem}
\usepackage{hyperref}
\usepackage{slashed}
\usepackage{amsmath}
\usepackage{epsfig}
\usepackage{amsfonts}
\usepackage{amssymb}
\usepackage{ulem}
\usepackage{wrapfig}

\DeclareMathOperator\arctanh{arctanh}

\def\beq{\begin{equation}}
\def\eeq{\end{equation}}
\def\bea{\begin{eqnarray}}
\def\eea{\end{eqnarray}}
\def\nn{\nonumber}
\def\tr{\text{Tr}}

\title{\boldmath The Unruh Quantum Otto Engine}
\author[a]{Enrique Arias,}
\author[b]{Thiago R. de Oliveira,}
\author[b]{M. S. Sarandy}
\affiliation[a]{Instituto Polit\'ecnico, Universidade do Estado do Rio de Janeiro, 28625-570 Nova Friburgo, Brazil.}
\affiliation[b]{Instituto de F\'{\i}sica, Universidade Federal Fluminense, 24210-346 Niter\'oi, Brazil.}

\emailAdd{earias@iprj.uerj.br}
\emailAdd{tro@if.uff.br}
\emailAdd{msarandy@id.uff.br}

\abstract{
We introduce a quantum heat engine performing an Otto cycle by using the thermal properties 
of the quantum vacuum. Since Hawking and Unruh, it has been established that the vacuum 
space, either near a black hole or for an accelerated observer, behaves as a bath of thermal 
radiation. In this work, we present a fully quantum Otto cycle, which relies on 
the Unruh effect for a single quantum bit (qubit) in contact with quantum vacuum fluctuations. 
By using the notions of quantum thermodynamics and perturbation theory  
we obtain that the quantum vacuum can exchange heat and produce work on the 
qubit. Moreover, we obtain the efficiency and derive the conditions to have both a thermodynamic 
and a kinematic cycle in terms of the initial populations of the excited state, which define
a range of allowed accelerations for the Unruh engine. 
}


\date{}
\begin{document}

\maketitle

\newpage
\section{Introduction}

There has been renewed interest in the investigation of the thermodynamics of quantum systems, 
namely, quantum thermodynamics~\cite{Gemmer:book}. A great deal of effort has been devoted to the 
foundations of the theory, with general discussions about the derivation of the second law 
in the quantum realm or even the whole theory of thermodynamics from 
quantum information theory~\cite{Goold16}. There is also intense focus on quantum heat engines, 
Maxwell demons, and advantages of work extraction by global operations over quantum 
systems~\cite{Maruyama:09}. Even though the first works in this direction appeared already 
in the 1950's \cite{Scovil59}, it regained interest through the works by Kieu~\cite{Kieu04,Kieu06}, 
where a quantum heat engine is proposed by using as working substance a 
single two-level system, i.e, a quantum bit (qubit). There, it has been analyzed the quantum 
Otto cycle, which is the quantum generalization of the classical prototype model of combustion 
engines. The Otto cycle is composed of two adiabatic and two isochoric steps operating between 
a hot and a cold reservoir. In order for the machine to produce a finite amount of work, one needs 
to fulfill the condition $T_H>(\Delta_2/\Delta_1)T_C$, with $T_H$ ($T_C$) the temperature of the 
hot (cold) reservoir and $\Delta_1$ and $\Delta_2$ the energy gaps in the qubit Hamiltonian 
spectrum before and after the adiabatic expansion, respectively. Therefore, for quantum systems, 
the classical inequality $T_H>T_C$ has been shown to be insufficient for a heat engine to produce 
work in individual thermodynamic cycles. Another possibility considered in Refs.~\cite{Kieu04,Kieu06} 
 is to use a gas of bosons as the thermal reservoir and let the interaction with the system to last a 
finite time, such that the system does not fully equilibrate with the reservoir. In that case,  a cycle 
can be built by suitably adjusting the initial population of the excited state.These works have been 
generalized to include different models for the reservoir and also for more complex working 
substances, e.g. coupled qubits, where quantum correlations such as entanglement may  
improve the thermal efficiency~\cite{Zhang:07,Zhang:08,Wang:09}. 

On the other hand, the relation between quantum thermodynamics, relativistic quantum mechanics, 
and black hole physics remains little explored. The standard thermodynamics of black holes 
has been elucidated a long time ago \cite{bardeen1973}, with striking results such as 
the Bekenstein-Hawking entropy area law~\cite{bekenstein1973} and Hawking 
radiation~\cite{hawking1974,hawking1975,hawking1976}. In this scenario, relativistic effects 
in quantum thermodynamics may lead to fruitful directions. A particularly relevant phenomenon is 
the Unruh effect and its implications to the domain of quantum thermodynamics.
The Unruh effect predicts the existence of thermal properties to the vacuum state of a quantum field. 
More specifically, it establishes that the quantum vacuum fluctuations with respect to an accelerated 
observer in flat space-time exhibit thermal properties with temperature proportional to the observer 
acceleration \cite{birrell, unruh1976}. Even though an experimental realization of the Unruh effect 
remains as a challenge, it has recently appeared a number of proposals of physical 
observations, such as in superconducting qubits~\cite{Felicetti:15}, ion traps~\cite{Laguna:17}, 
quantum metrology~\cite{Wang:14}, oscillating neutrinos~\cite{Ahluwalia:16}, and classical 
electrodynamics~\cite{Cozzella:17} (see also Ref.~\cite{Crispino:08} for other previous proposals). 
Here, our aim is to exploit the thermal properties of the vacuum fluctuations predicted by the Unruh effect 
to model a thermal reservoir and define a relativistic quantum heat engine. We consider as working 
substance a single qubit linearly coupled to a quantum field prepared in its vacuum state. This is the 
model used by Unruh and DeWitt in order to detect the quantum vacuum fluctuations 
effects~\cite{unruh1976, dewitt1979}. The Unruh quantum thermodynamical cycle proposed in our 
work consists of a series of quantum adiabatic and isochoric processes. During the adiabatic processes, 
we consider that the qubit undergoes an expansion/contraction of its energy gap due to the influence of 
some external field. Since the processes are adiabatic, there is no heat exchange between the qubit and 
the environment. However, the adiabatic expansion/contraction would cost/produce some amount of work. 
On the other hand, the isochoric processes are defined here by coupling the qubit to the quantum vacuum 
and considering two different constant accelerations for the qubit at these stages. In this manner, due to 
the Unruh effect, the qubit would feel the vacuum as a thermal reservoir, with no work at this stage.
Heat and work are provided by the prescriptions of quantum thermodynamics 
and are find using a perturbation theory approach in the coupling constant between the qubit 
and the quantum field. 
In order to run as a thermal machine, we must also consider the conditions for a 
closed trajectory of the qubit in real space, since we are dealing with simultaneously thermodynamic 
and kinematic cycles. The cycle conditions are then derived  in terms of the initial populations of the 
excited state, which define a range of allowed acceleration pairs for the Unruh quantum engine. 
Throughout the paper, we will adopt natural units, such that  the speed of light $c$, the Planck constant $\hbar$, 
and the Boltzmann constant $k_B$ are taken as $c=\hbar=k_B=1$.

\section{Quantum Otto engine}

In this section, we briefly review some notions of quantum thermodynamics and quantum heat 
engines. The basic definitions here are those of work and heat in a quantum process. For a quantum 
system described by a density operator given by $\rho(t)$ evolving under a time-dependent Hamiltonian 
${\cal H}(t)$, the energy expectation value $ \langle E(t) \rangle = \tr (\rho(t){\cal H}(t))$ satisfies
\begin{equation}
\partial_t \langle E(t) \rangle=\tr (\partial_t \rho(t){\cal H}(t)) + \tr (\rho(t)\partial_t{\cal H}(t)).
\label{energy-der}
\end{equation}
We see that the first term on the right hand side of Eq.~(\ref{energy-der}) is associated with changes 
in the inner state of the system, which are related to the populations of $\rho(t)$, 
while the second term has its origin only in changes of the external parameters in ${\cal H}(t)$. 
We thus identify the first as the heat transfer $\langle Q \rangle$ and the second as the work $\langle W \rangle$ 
realized on the system. This association provides 
\beq
\langle Q \rangle=\int_0^{\cal T}dt\,\,\tr\left(\frac{d\rho(t)}{dt}{\cal H}(t)\right),
\eeq
and 
\beq
\langle W \rangle=\int_0^{\cal T}dt\,\,\tr\left(\rho(t)\frac{d{\cal H}(t)}{dt}\right),
\eeq
where ${\cal T}$ denotes the total time of evolution and quasi-static processes are adopted. 
For the processes of interest here, either $\langle Q \rangle$ or $\langle W \rangle$ will be 
vanishing. Then, from the first law of thermodynamics, this will allow to take these quantities as 
dependent only on the initial final state of the system. More general cases must be carefully 
examined (for a review see, e.g., Ref.~\cite{Campisi:11}).  
It is worth observing here that the notion of heat in the quantum thermodynamics domain
is entirely compatible with the notions of Boltzmann entropy $S=-\tr\left(\rho\ln\rho\right)$
and that of Boltzmann distribution at equilibrium $\rho=e^{-{\cal H}/T}/Z$, where $Z$ is a 
normalization constant and $T$ is the temperature of the system. 
By using the definition of heat $\delta Q=TdS$, it is straightforward to show that 
$\delta Q=\tr(d\rho\,{\cal H})$, which is equivalent to the definition of mean quantum heat given above.

We now briefly review the Otto quantum engine first proposed in Ref.~\cite{Kieu04}. 
The classical Otto cycle is composed of two adiabatic processes together with two isochoric 
processes, which is common in automobile piston engines. In its quantum version, 
the working substance is a quantum system instead of an ideal gas. Here, we adopt the simplest 
quantum system, namely, a qubit. The two adiabatic processes, which in the classical engine are 
provided by the adiabatic expansion and contraction of a piston, here are adiabatic increase and 
decrease of the upper energy level of the qubit, respectively. The two isochoric process of the 
classical case are replaced in the quantum version  by keeping the system with constant gaps 
while in contact with a thermal reservoir, which implies in vanishing work. 

Let us now consider a single qubit with eigenlevels $\{|e\rangle,|g\rangle\}$, which are associated 
with energies $\{\omega, 0\}$, respectively. Hence, by applying the spectral theorem~\cite{Horn:13}, 
the qubit Hamiltonian can be written as ${\cal H}=\omega\, |e\rangle\langle e|$. 
In our cycle we consider that, during an adiabatic process, the energy level $\omega$ changes.
This change could be caused by some external field or changes in the external conditions as the
size of the recipient where the qubit is confined. Here, we only assume it can be done
in a smooth way such that the quantum adiabatic theorem is valid~\cite{Born:28,Kato:50,Messiah:book}.
The four stages of a quantum Otto cycle are 
\begin{itemize}

\item (i) The qubit begins at the initial state $\rho_{in}=p |e \rangle \langle e|+(1-p) |g \rangle \langle g|$
 with an energy gap $\omega_1$ and undergoes an adiabatic expansion to a larger value $\omega_2$.
This expansion is at the cost of some work over the system. On the other hand, since the process is adiabatic, 
no heat is exchanged with the environment. 

\item (ii) The qubit in the state $\rho_{in}$ and with fixed energy gap $\omega_2$ 
is put in contact with a hot bath at temperature $T_H$.
After some interaction time ${\cal T}_2$, the qubit ends up at the state 
$\rho= (p+\delta p_H) |e \rangle \langle e|+(1-p-\delta p_H) |g \rangle \langle g|$. 
Only heat is transfered in this process, with no work performed. 

\item (iii) The qubit at the state $\rho$ is isolated from the hot bath and undergoes an adiabatic 
compression, which reduces its energy gap from $\omega_2$ to $\omega_1$. This process is 
adiabatic, with no change in the populations of the energy levels. Thus, work is performed by the system, 
with no heat transfered. 

\item (iv) The qubit still with the state $\rho$ is put in contact with a cold reservoir 
at temperature $T_C$.  After some interaction time ${\cal T}_1$, the qubit ends up at the final state 
$\rho_{fin}= (p+\delta p_H+\delta p_C) |e \rangle \langle e|+(1-p-\delta p_H-\delta p_C) |g \rangle \langle g|$. 
Heat is transfered between the qubit and the environment, with no work performed. 
\end{itemize}

One of the novelties in the quantum regime is that the processes are now probabilistic, i.e., 
while we expect in the classical case that heat flows from the hot bath to the system for every 
single cycle at stage (ii), namely, $\delta p_H>0$, there is in the quantum case a finite probability, 
for individual cycles, of the opposite to occur, namely, $\delta p_H<0$. Thus, it is possible to observe 
a violation of the behavior expected by the second law of thermodynamics for individual quantum cycles. 
Naturally, conciliation with the second law is recovered on average over a long sequence of cycles~\cite{Kieu04}. 
Claims of violation of the second law have been considered, but they all typically use non-thermal baths 
(see, e.g., Ref.~\cite{Niedenzu17}). Note also that the random violation of the second law for individual 
quantum cycles are due to quantum fluctuations rather than to thermal fluctuations. 

The probabilistic character of a quantum engine implies that we need to enforce the cyclicity of the engine 
through the constraint that $\delta p_H + \delta p_C=0$, which ensures that the state of the system at the 
end of stage (iv) is equal to the initial state at the beginning of stage (i). By imposing such a constraint, 
we obtain that the total amount of work performed by the heat engine is
\beq
\Delta W = \delta p_H(\omega_2 - \omega_1).
\eeq
So it is clear that in order to have a positive work and therefore a heat engine, instead of a refrigerator, 
we need to require $\delta p_H>0$. 
When the system is let to thermalize with the bath, with $\tau_1,\tau_2 \rightarrow \infty$,
the state of the qubit at the end of stage (ii) is such that 
$p+\delta p_H = \tr \left[\, |e\rangle\langle e| \,\rho \, \right] = 1/ \left[1+\exp(\beta_H \omega_2) \right]$, 
where $\beta_H = 1/T_H$. Similarly at the end of stage (iv), 
as the system returns to its original state, we have 
$p = \tr \left[\, |e\rangle\langle e|\, \rho_{fin} \, \right] = 1/ \left[1+\exp(\beta_C \omega_1) \right]$, 
where $\beta_C = 1/T_C$. It then follows that the condition for positive work is, on average,  
$T_H> (\omega_2/\omega_1)T_C$. This is stronger than the classical condition of $T_H>T_C$.
It can be understood by the fact that the system is expected to receive heat from the hot bath if 
$T_H > \omega_2$ and to transfer heat to the cold bath if $T_C < \omega_1$.

\section{Quantum Otto engine via Unruh effect}

\begin{figure}[ht]
\centering
\includegraphics[scale=0.35]{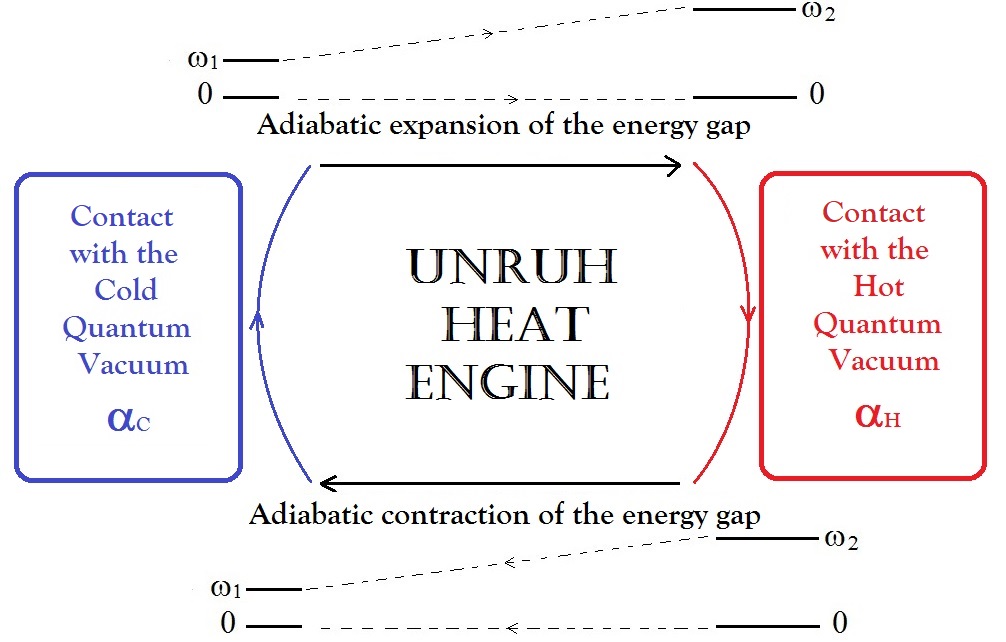}
\caption{The Unruh quantum thermodynamic cycle.}
\label{fig:unrhucycle}
\end{figure}

We now present the Unruh quantum Otto engine. In this case, the thermal bath comes from the 
thermal properties of the vacuum fluctuations given by the Unruh effect. Thus, the thermal reservoir 
has now a relativistic quantum origin, which is associated with the vacuum fluctuations of a quantum field. 
This implies that the Unruh effect is able to yield useful work in a cyclic process. As a working substance, 
we take a single qubit, which is designed to transform heat extracted from the vacuum into work. 
The stages proposed to the Unruh thermodynamic cycle are illustrated in Fig.~(\ref{fig:unrhucycle}).

\subsection{Kinematic cycle of the qubit}

\begin{figure}[ht]
\centering
\includegraphics[scale=0.5]{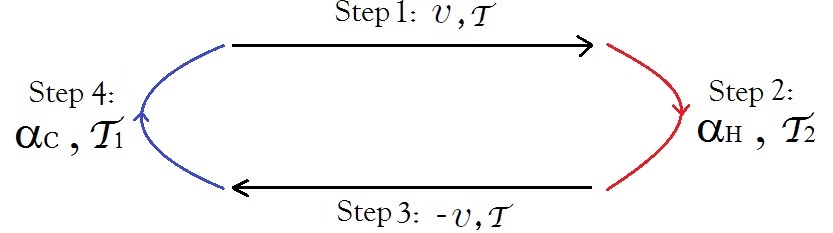}
\caption{The relativistic kinematic cycle of the qubit.}
\label{kine}
\end{figure}

Since the observation of the Unruh effect depends on the state of motion of the qubit as it interacs with 
the quantum vacuum, a description of the kinematics of the qubit on the whole cycle is required.
In this way, in addition to a thermodynamic closed cycle, we also assume a kinematic closed trajectory 
for the qubit traveling in spacetime. This will impose further constraints in the interaction time with the 
quantum vacuum, as it will be shown below. Thus, in the first and third step of the cycle, we assume the
qubit keeps constant speed , while in the second and fourth steps, it moves with constant acceleration. 
This is explicitly shown in Fig.~(\ref{kine}). Notice that, while traveling at constant acceleration the qubit 
is brought into contact with the vacuum fluctuations. On the other hand, while at constant velocity and 
going through an adiabatic expansion/contraction of the energy gap, there is no contact between the 
qubit and the quantum field.

In order to assure that there is no change in the state of the system, caused by the vacuum, during 
the adiabatic processes of the cycle, we should keep the qubit decoupled of the vacuum at these steps. 
We are thus assuming that one can control the contact between the system and the vacuum in the very 
same way that one can control the coupling or decoupling between a system and a thermal reservoir in 
a classical thermal machine. For simplicity we assume a one-dimensional motion of the qubit.
Thus, our closed trajectory is composed by the following steps:
\begin{itemize}
\item \textit{Step} 1: Constant velocity $v$ during a period of time ${\cal T}$.	
\item \textit{Step} 2: Constant acceleration $\alpha_H$ during a period ${\cal T}_2$. At this step, 
the qubit accelerates from speed $v$ to $-v$. 
\item \textit{Step} 3: Constant velocity $-v$ during a period of time ${\cal T}$.
\item \textit{Step} 4: Constant acceleration $\alpha_C$ during a period ${\cal T}_1$.
At this step the qubit accelerates from speed $-v$ to $v$. 
\end{itemize}
 It is worth noticing that times ${\cal T}_{1}$ and ${\cal T}_{2}$ are fixed by the values of $v$ and $\alpha_{H,C}$.
More specifically, we will shown that there is a relation between $\alpha_H$ and $\alpha_C$ in order to ensure 
that the system returns to its original state after one cycle. For now, we say that it is expected the accelerations 
(temperatures) satisfy $\alpha_H>\alpha_C$ to ensure that at step 2 the vacuum behaves like a  hot reservoir 
while at step 4 it acts like a cold reservoir.

\begin{figure}[ht]
\centering
\includegraphics[scale=0.3625]{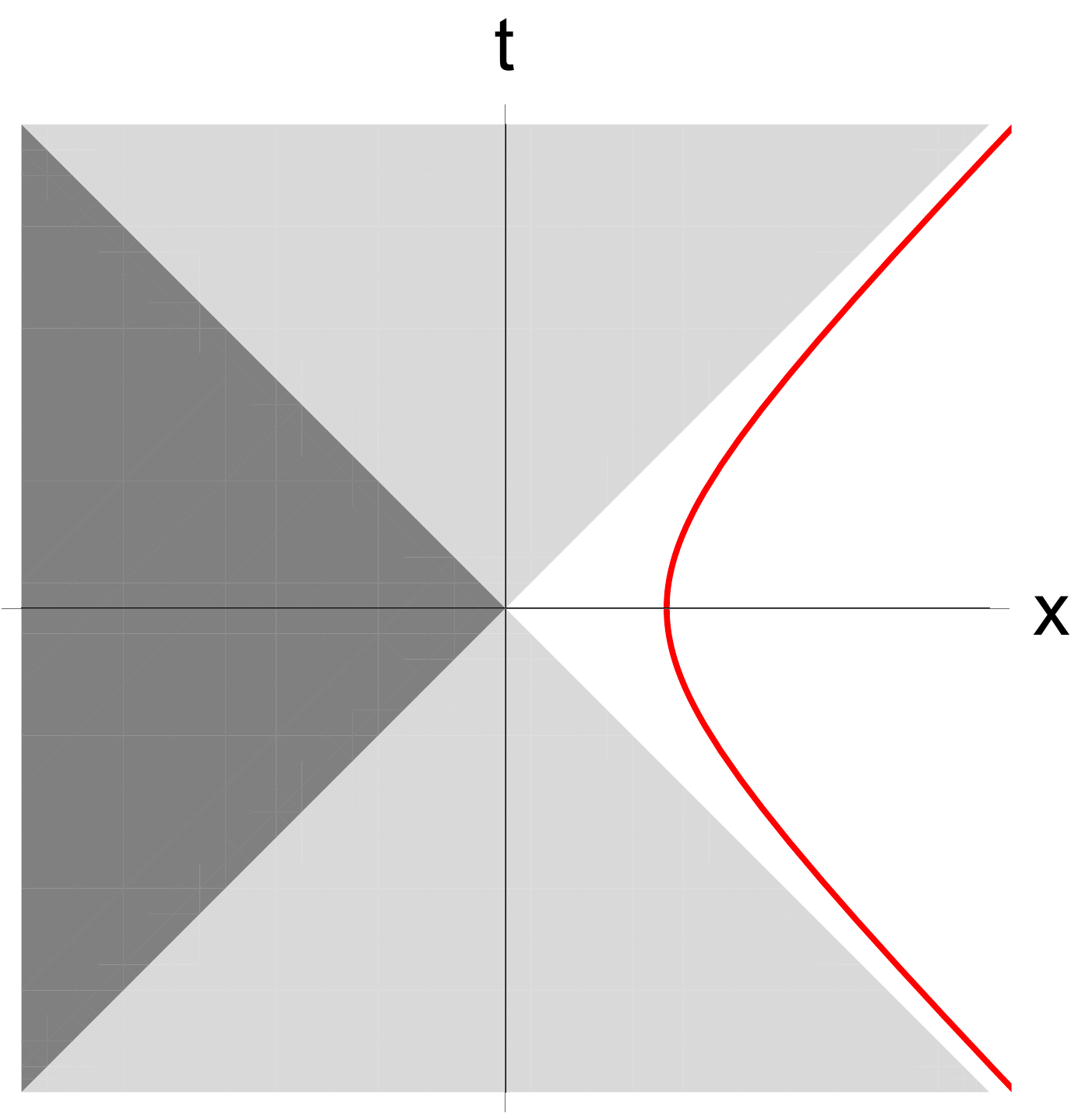}
\caption{Hyperbolic trajectory of a accelerated qubit. 
In light gray are shown spacetime regions partially causally disconnected from the qubit.
In dark gray is the region totally causally disconnected from the qubit.
The white region is the causally accesible region of the qubit, the border of this region represents
an event horizon for the accelerated qubit.}
\label{hiperbola}
\end{figure}

The non-trivial motion of the qubit occurs when it is in contact with the vacuum. Indeed, in order for the vacuum 
to behave as a thermal bath, we have to ensure a motion with constant proper acceleration for the 
qubit~\cite{birrell}. Therefore, during the stages of contact with the vacuum, we suppose that the qubit 
moves with constant proper acceleration, as described by the step 2 and step 4 of the kinematics cycle. 
This would ensure the thermal behavior of quantum vacuum as predicted by the Unruh effect. 
In general, if one considers a relativistic motion with constant proper acceleration $\alpha$,
one finds that the spacetime coordinates of the particle are given by $\chi(\tau)=(t,x)$ where
\bea
t&=&\frac{1}{\alpha}\sinh(\alpha\tau),\nn\\
x&=&\frac{1}{\alpha}\cosh(\alpha\tau),
\label{tx}
\eea
being $\tau$ the particle proper time. This hyperbolic spacetime trajectory 
of the qubit with constant acceleration is shown in Fig.~(\ref{hiperbola}).
From Eqs.~(\ref{tx}), the velocity of the qubit is given by
\beq
v=\tanh(\alpha\tau).
\eeq
Then, we see then that if the qubit velocity begins at $-v$ at time $-\tau$, it will have
velocity $v$ at time $\tau$, where 
\beq
\tau=\frac{1}{\alpha}\arctanh(v).
\eeq
Then, the time that it takes for the qubit to change from velocity $-v$ to $v$ using a constant acceleration 
$\alpha$ is $2\tau=2\arctanh(v)/\alpha$. Hence, we have that, when the qubit has acceleration $\alpha_H$, 
the interaction time with the vacuum is ${\cal T}_2=2\arctanh(v)/\alpha_H$, whereas when the qubit has 
acceleration $\alpha_C$ the interaction time with the vacuum is ${\cal T}_1=2\arctanh(v)/\alpha_C$.

\subsection{Adiabatic expansion of the energy gap}

Now we describe the first stage of our engine. 
This process corresponds to an adiabatic expansion of the energy gap, 
where we assume that the energy gap of the qubit increases smoothly in time. 
Recalling that the qubit Hamiltonian is given by ${\cal H}=\omega(t)|e\rangle\langle e|$ 
and that we consider as the initial state of the qubit the density operator
$\rho_{in}=p |e \rangle \langle e|+(1-p) |g \rangle \langle g|$, it follows that the 
state of the qubit all along this adiabatic process is kept constant, so $\rho(t)=\rho_{in}$.
Therefore, we obtain that during the adiabatic expansion there is no heat absorbed by the qubit, i.e.
\beq
\langle Q_1\rangle=\int_0^{\cal T}dt\,\,\tr\left(\frac{d\rho(t)}{dt}{\cal H}(t)\right)=0.
\eeq
However the energy gap expansion of the qubit comes at the cost of work on it, which reads
\bea
\langle W_1\rangle&=&\int_0^{\cal T}dt\,\,\tr\left(\rho(t)\frac{d{\cal H}(t)}{dt}\right),\nn\\
&=&\int_0^{\cal T}dt\,\,\tr\left(\rho_{in}\frac{d(\omega(t)|e\rangle\langle e|)}{dt}\right),\nn\\
&=&\int_0^{\cal T}dt\,\,\frac{d\omega(t)}{dt}\,\,\tr\left(\rho_{in}|e\rangle\langle e|\right).
\eea
We now use that $\tr\left(\rho_{in}|e\rangle\langle e|\right)=p$, where $p$ is the probability of finding 
the system in the excited state $|e\rangle$. Moreover, considering that the excited state increases its 
energy level in the adiabatic expansion from $\omega(t=0)=\omega_1$ to $\omega(t={\cal T})=\omega_2$, 
we get a positive work given by
\beq
\langle W_1\rangle=p\,\left(\omega_2-\omega_1\right).
\eeq

\subsection{Contact with the hot quantum vacuum}

In this part of the cycle we take advantage of the Unruh effect to simulate a thermal bath to the qubit.
We take the qubit as interacting with a quantum massless scalar field in its ground state of zero 
particles, i.e. in its vacuum state $|0\rangle$. Due to the Unruh effect we know that this 
quantum vacuum would behaved like a thermal bath with respect to the qubit when it
moves in an accelerated trajectory. Therefore the kinematics description of the qubit 
previously presented is crucial for this part of the cycle. 
An important remark here is that we suppose that
the qubit system is small enough so that it does not affect the 
state of the scalar field and disturb the quantum vacuum state.

During the interaction of the qubit with the quantum scalar field, 
we consider that the Hamiltonian of the total qubit-field system 
is given by
\beq
\mathbb{H}=\mathbb{H}_0+\mathbb{H}_{int},
\eeq
where $\mathbb{H}_0$ is the free Hamiltonian and $\mathbb{H}_{int}$ is the 
qubit-field interaction Hamiltonian. The Hamiltonian $\mathbb{H}_0$ reads
\beq
\mathbb{H}_0={\cal H}+{\cal H}^{field},
\eeq
where ${\cal H}$ denotes the qubit free Hamiltonian  ${\cal H}=\omega_2 |e\rangle\langle e|$, 
with $\omega_2$ the constant gap along this stage, and 
${\cal H}^{field}=\int d^3x (1/2)\{(\partial_t\varphi)^2+(\nabla\varphi)^2\}$
is the Klein-Gordon Hamiltonian associated to a free massless scalar field $\varphi$. 
The qubit-field interaction Hamiltonian
is given by the linear coupling \cite{dewitt1979}
\beq
\mathbb{H}_{int}= g\,m\varphi(\chi(\tau)),
\eeq
where $g$ is the coupling constant of the interaction,
$m$ is the qubit monopole operator, which is given by
\beq
m=|e\rangle\langle g|+|g\rangle\langle e|,
\eeq
and the scalar field $\varphi(\chi(\tau))$ is evaluated on the spacetime point $\chi(\tau)=(t,x)$
where the qubit is located.
We suppose at this stage that the qubit has a constant acceleration $\alpha_H$,  
with a spacetime trajectory analog to that given in Eq. (\ref{tx}). 

We consider that the scalar field start its evolution at the vacuum state
and the initial state of the qubit-field is the tensor product  
$\varrho_{in}=\rho_{in}\otimes|0\rangle\langle0|$.
We adopt here the interaction picture, where we apply a perturbative approach  
and a Dyson series to obtain the final state of the qubit up to second order in the 
small coupling constant $g$. This is shown in details in Appendix \ref{ap:dynamics}. 
We then show that the final state of the qubit after interacting with the vacuum is
\beq
\rho(t)=\rho_{in}+\delta p_H(t)\sigma_3,
\eeq
where $\sigma_3 = (|e\rangle\langle e | - |g\rangle\langle g |)$ is the third Pauli matrix and 
the time $t$ during this vacuum contact is restricted to
$-\tau_2<t<\tau_2$,
with $\tau_2=\arctanh(v)/\alpha_H$.
The change in the population of the excited state is given by
\beq
\delta p_H(t)=g^2\int_{-\tau_2}^td\tau\int_{-\tau_2}^td\tau'\left((1-p)e^{-i\omega_2\Delta\tau}-p\,e^{i\omega_2\Delta\tau}\right)
G^+_{\alpha_H}(\tau,\tau'),
\label{deltappp}
\eeq
where $\Delta\tau=\tau-\tau'$ and $G^+_{\alpha_H}(\tau,\tau')$ is the Green correlation function 
of the scalar field evaluated at two points of the qubit accelerated trajectory.
This vacuum correlation function evaluated on the accelerated trajectory of the qubit
captures the thermal nature of the vacuum fluctuations.
Therefore, at some time $t$ during the interaction with the quantum vacuum the density operator of the 
qubit is given by
\beq
\rho(t)=\left(
\begin{array}{cc}
p+\delta p_H(t) & 0\\
0 & 1-(p+\delta p_H(t))
\end{array}
\right).
\label{state}
\eeq
One can see that the increase of the population of the excited state, Eq. (\ref{deltappp}), 
is caused by the interaction with the vacuum and this would give us a thermal like response,
as predicted by the Unruh effect. We also observe that the change in the probability of 
finding the qubit in its excited state depends on the initial population of this state $p$, 
on the energy gap $\omega_2$ and on the acceleration $\alpha_H$, which plays the role 
of \textit{hot temperature} in the Unruh quantum machine.

We can then evaluate the average heat and work over the qubit
during the contact with the hot quantum vacuum. From the state of the qubit given 
by Eq. (\ref{state}), at any instant in the interval of contact $-\tau_2<t<\tau_2$, we obtain 
the  mean quantum heat absorbed from the quantum vacuum 
\beq
\langle Q_2\rangle=\int_{-\tau_2}^{\tau_2}dt\,\,\tr\left(\frac{d\rho(t)}{dt}{\cal H}(t)\right).
\eeq
Since 
at this part of the cycle the free qubit Hamiltonian is 
${\cal H}=\omega_2\,|e\rangle\langle e|$, 
the mean quantum heat extracted from the vacuum is
\bea
\langle Q_2\rangle&=&\int_{-\tau_2}^{\tau_2}dt\,\,\tr\left(\frac{d\rho(t)}{dt}{\cal H}\right),\nn\\
&=&\int_{-\tau_2}^{\tau_2}dt\,\,\tr\left(\frac{d(\delta p_H(t)\sigma_3)}{dt}\,\omega_2|e\rangle\langle e|\right),\nn\\
&=&\omega_2\left(\int_{-\tau_2}^{\tau_2}dt\,\,\frac{d(\delta p_H(t))}{dt}\right)\tr(\sigma_3|e\rangle\langle e|),\nn\\
&=&\omega_2\left(\delta p_H(\tau_2)-\delta p_H(-\tau_2)\right),\nn\\
&=&\omega_2\,\delta p_H,
\label{q2}
\eea
where we have used that, from Eq. (\ref{deltappp}), we have $\delta p_H(-\tau_2)=0$ and 
we have also defined the total correction $\delta p_H=\delta p_H(\tau_2)$, which is given by 
\beq
\delta p_H
=g^2\int_{-\tau_2}^{\tau_2}d\tau\int_{-\tau_2}^{\tau_2}d\tau'\left((1-p)e^{-i\omega_2\Delta\tau}-p\,e^{i\omega_2\Delta\tau}\right)
G^+_{\alpha_H}(\tau,\tau').
\label{deltaph}
\eeq
Note that $\langle Q_2\rangle$ depends on the energy gap $\omega_2$ and 
on the change $\delta p_H$ of the excited level population due to the hot vacuum fluctuations. 
We will discuss ahead the conditions to ensure that $\delta p_H$ 
and the heat absorbed from the vacuum $\langle Q_2\rangle$ are both positive, 
so the quantum vacuum here really behaves as a hot reservoir.
As at this stage of the cycle, the qubit Hamiltonian is constant  ${\cal H}=\omega_2|e\rangle\langle e|$,
which implies that we get no work, i.e.
\beq
\langle W_2\rangle=\int_{-\tau_2}^{\tau_2}dt\,\,\tr\left(\rho(t)\frac{d{\cal H}}{dt}\right)=0.
\eeq

\subsection{Adiabatic contraction of the energy gap}

We suppose next that the energy gap is contracted by means of external fields.
The specific way this contraction occurs is not important, but in order to ensure 
a closed cycle for the quantum heat engine, we have to get back to the initial energy gap. 
Then, at this stage we decouple the qubit from the vacuum. Hence, the system is kept in the state
\beq
\rho=\left(
\begin{array}{cc}
p+\delta p_H & 0\\
0 & 1-(p+\delta p_H)
\end{array}
\right).
\label{rhoa}
\eeq
Again, the change in the
energy levels of the qubit Hamiltonian does not affect the state of the qubit.
Then, the state of the qubit remains fixed as in Eq. (\ref{rhoa}) all over the
adiabatic contraction and there is no heat transfer.
On the other hand, the qubit realizes work, which is proportional to the 
energy gap change and to the population of the excited state. 
In this way we straightforwardly obtain that, during the adiabatic contraction, heat 
and work read
\bea
\langle Q_3\rangle&=&0,\nn\\
\langle W_3\rangle&=&(p+\delta p_H)(\omega_1-\omega_2),
\eea
where the energy gap contraction is taken from $\omega_2$ to the initial lower value 
$\omega_1$ considered at stage 1 of the quantum cycle.

\subsection{Contact with the cold quantum vacuum}

Here we consider that the qubit is bring into contact with the quantum vacuum again, 
but now with acceleration $\alpha_C <  \alpha_H$. So, at this stage, the vacuum plays the 
role of a the cold bath reservoir. 
The qubit has a lower energy gap $\omega_1$ in this process and its initial state is
\beq
\rho=\left(
\begin{array}{cc}
p' & 0 \\
0 & 1-p'
\end{array}
\right),
\eeq
where we have defined $p'=p+\delta p_H$. Then we have that the final state
of the qubit is given by
\beq
\rho_{fin}=\left(
\begin{array}{cc}
p'+\delta p_C & 0 \\
0 & 1-(p'+\delta p_C)
\end{array}
\right),
\label	{rhofinal2}
\eeq
where the new change in the probability of the excited state is given by 
\beq
\delta  p_C=g^2\int_{-\tau_1}^{\tau_1}d\tau\int_{-\tau_1}^{\tau_1}d\tau'\left((1-p')e^{-i\omega_1\Delta\tau}-p'\,e^{i\omega_1\Delta\tau}\right)G^+_{\alpha_C}(\tau,\tau'),
\label	{rhofinal}
\eeq
as discussed in Appendix~ \ref{ap:dynamics}. By considering only terms up to second order in the 
coupling constant and realizing that the first correction $\delta p_H$ of Eq. (\ref{deltaph}) is already 
of order two, we can show that
\beq
\delta  p_C=g^2\int_{-\tau_1}^{\tau_1}d\tau\int_{-\tau_1}^{\tau_1}d\tau'\left((1-p)e^{-i\omega_1\Delta\tau}-p\,e^{i\omega_1\Delta\tau}\right)G^+_{\alpha_C}(\tau,\tau').
\label{dpc}
\eeq
Note that here we have considered that the energy gap of the qubit is $\omega_1$ and that
the acceleration in this stage of the cycle is lower that previously $\alpha_C<\alpha_H$. 
This expression is very similar to that one we have obtained for the case of a hot quantum 
vacuum. However, in the same way that we must ensure that, during the first contact of the 
qubit with the quantum vacuum, it absorbed heat, i.e. $\delta p_H>0$, here we must require 
that, in order for the vacuum to behave as a cold reservoir, the qubit have to transfer heat to 
the vacuum, i.e., $\delta p_C<0$. As we will see, these conditions are not independent of each 
other, since we also need to ensure that during a cycle the qubit comes back to its original state. 
As before, we can obtain the average quantum heat and the work produced in this part of the cycle, 
which read
\bea
\langle Q_4\rangle&=&\omega_1\delta p_C,\nn\\
\langle W_4\rangle&=&0.
\eea
After this stage of the cycle, we must impose that the system comes back to its initial state. 
Then the final state of the qubit, given by Eq. (\ref{rhofinal2}), requires $\delta p_C+\delta p_H=0$. 
This cyclicity condition, together with the efficiency of the cycle, are analyzed in the next sections.

\section{Quantum vacuum fluctuations effects }

From our previous results, we have found that the corrections induced by the vacuum 
fluctuations to the qubit excited state population can be written in a general form as
\beq
\delta  p=g^2\int_{-\tau}^{\tau}dt \int_{-\tau}^{\tau}dt^{\prime}\left((1-p)e^{-i\omega\Delta t}-p\,e^{i\omega\Delta t}\right)G^+_{\alpha}(t,t^{\prime}),
\label{integral}
\eeq
where $\alpha$ is the qubit acceleration, $\omega$ is the qubit
energy gap, $p$ is the initial population of the excited state, and  $2\tau$ is the interaction 
time of the qubit with the vacuum. However, as we need to ensure a kinematically closed cycle 
for the qubit, we require that the interaction time of the qubit with the vacuum is such that 
\begin{equation}
\tau=\arctanh(v)/\alpha.
\label{tva}
\end{equation}
On the other hand, as it is shown in Appendix~ \ref{ap:regularization}, 
the qubit acceleration and the qubit energy gap appear as a single variable, which motivates 
the definition of the ratio $a\equiv\alpha/\omega$, here called as reduced acceleration. In turn, 
the correction to the excited state population Eq. (\ref{B18}), is then written as 
\beq
\delta p=\delta p(a,p,v) = g^2\left((1-2p)J\left(-\frac{1}{a},2\arctanh{v}\right)-
p\frac{\arctanh{v}}{2a}\right).
\label{delta}
\eeq

\begin{figure}[ht]
	\begin{subfigure}[b]{0.46\textwidth}
		\fbox{\includegraphics[width=\textwidth]{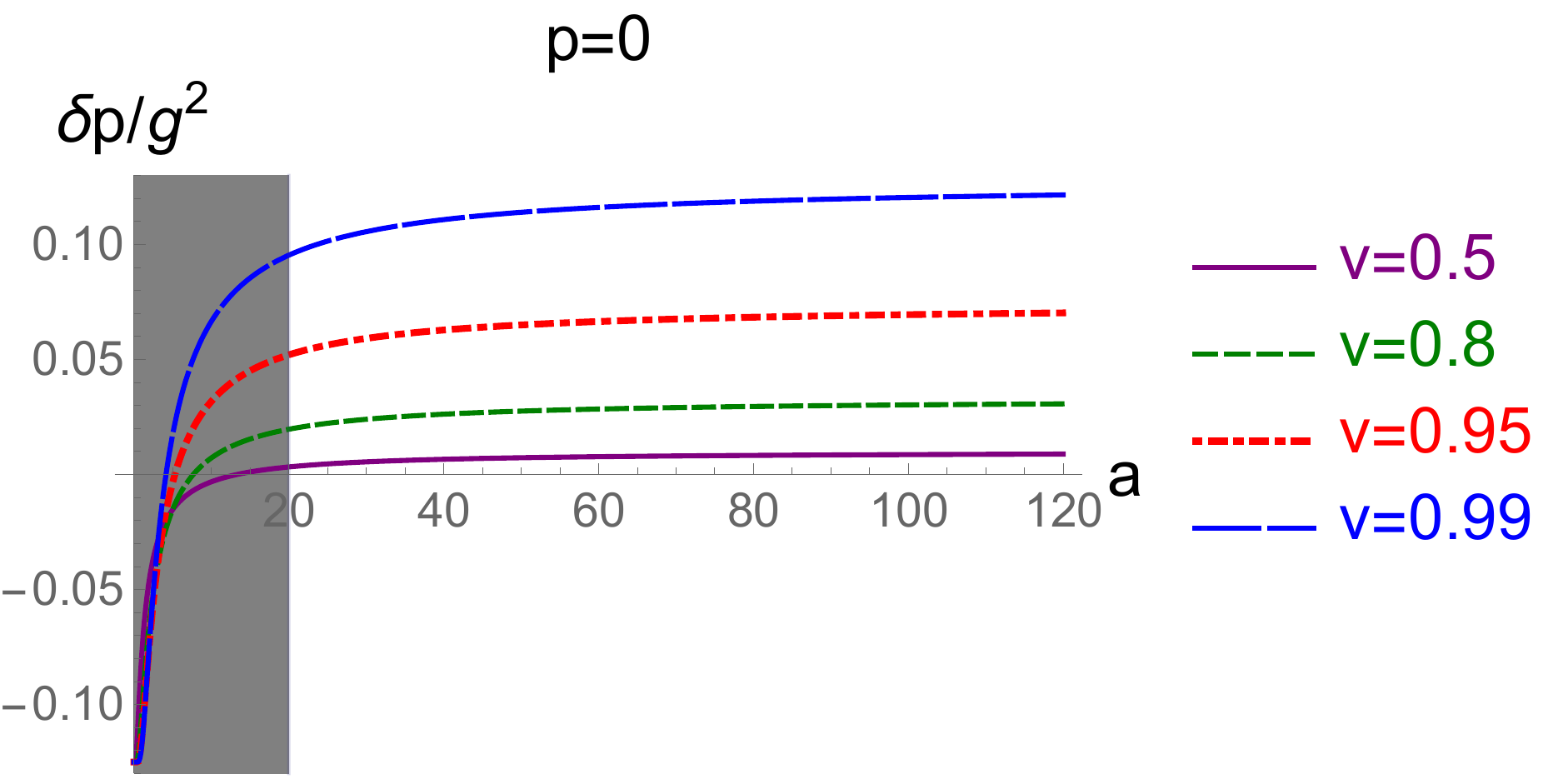}}
		\caption{Completely empty excited state.}
		\label{5a}
	\end{subfigure}
	\hfill
	\begin{subfigure}[b]{0.45\textwidth}
		\fbox{\includegraphics[width=\textwidth]{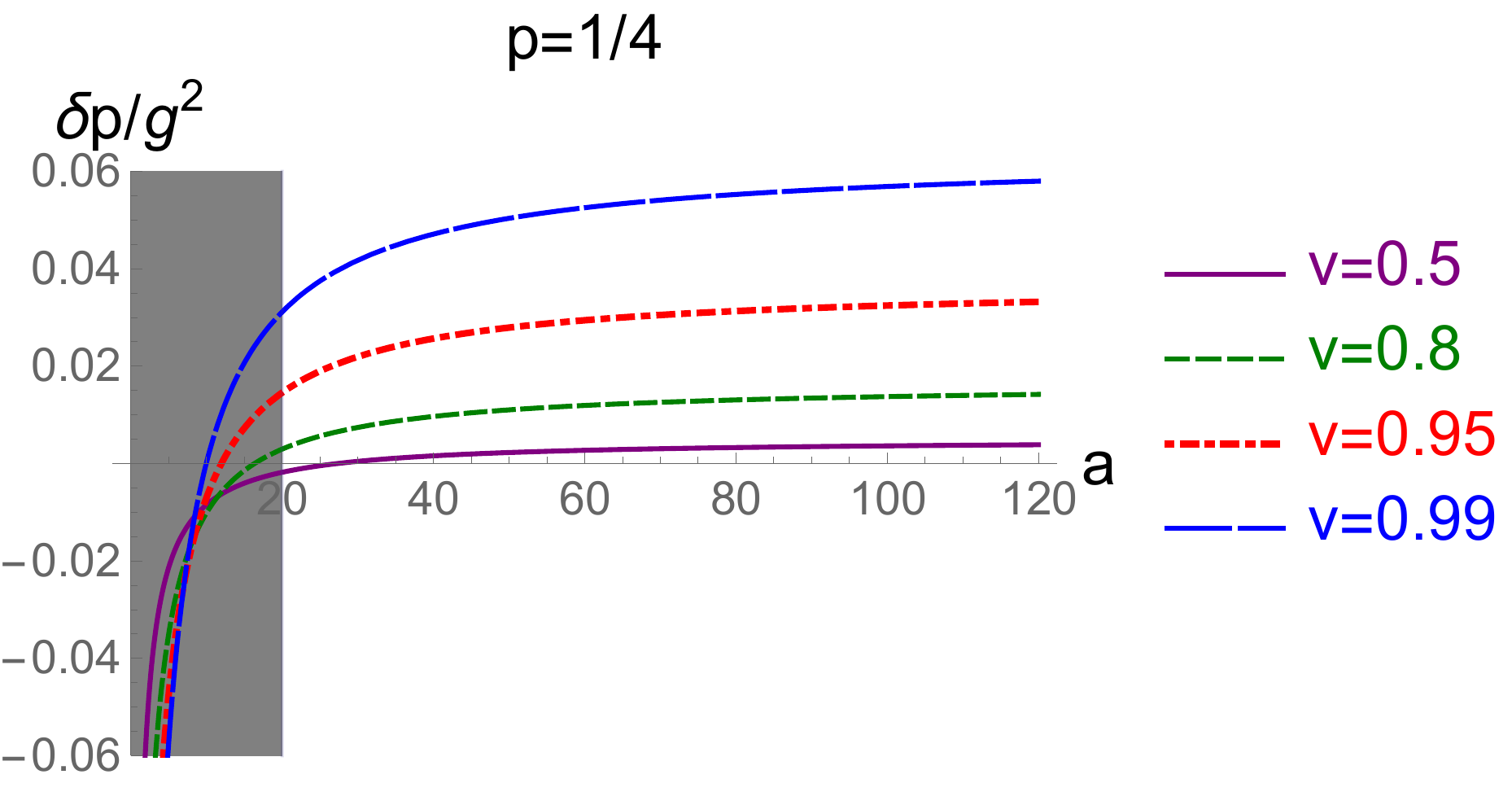}}
		\caption{Lowly populated excite state.}
		\label{}
	\end{subfigure}
	\hfill
	\begin{subfigure}[b]{0.46\textwidth}
		\fbox{\includegraphics[width=\textwidth]{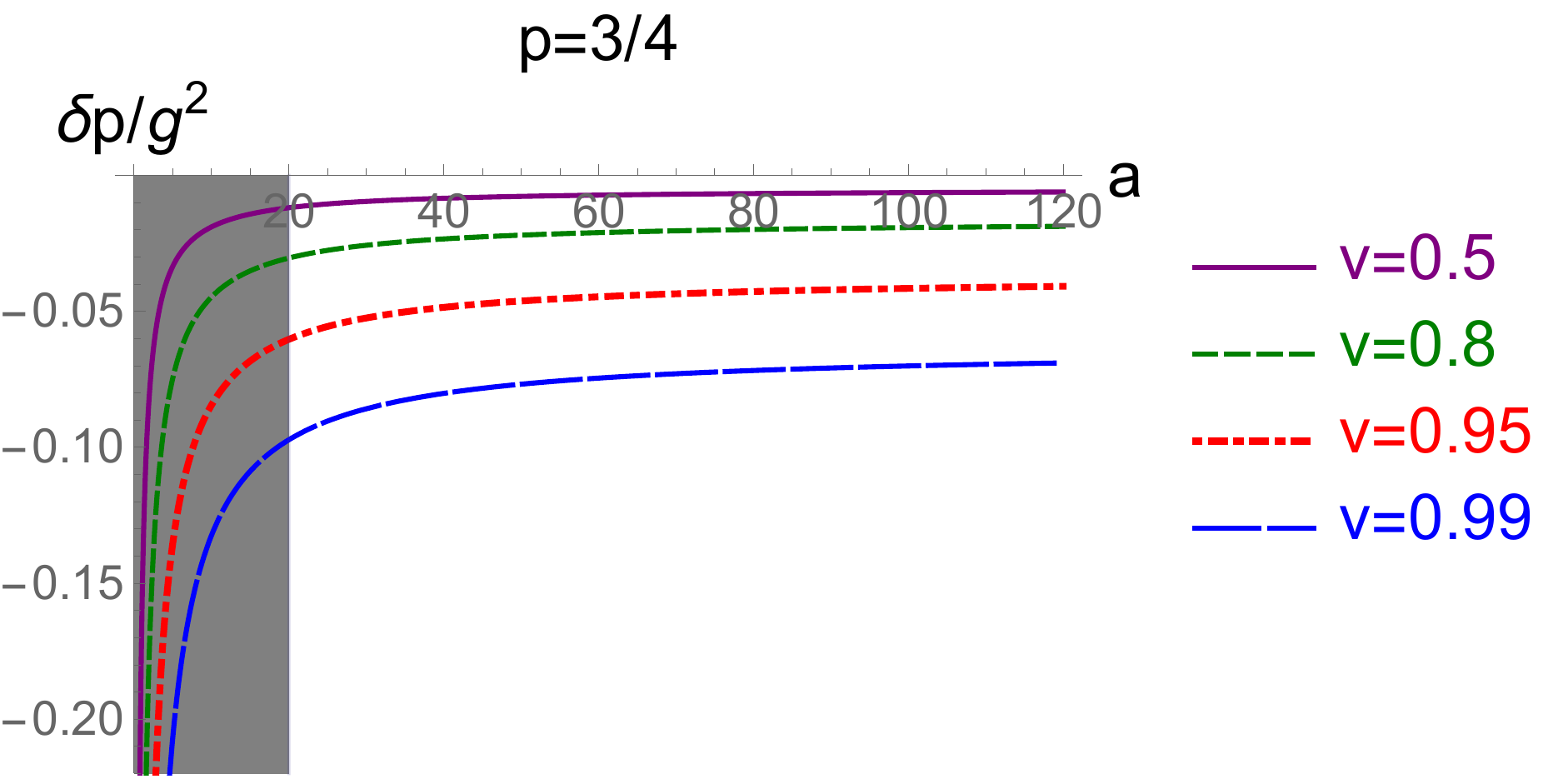}}
		\caption{Highly populated excited state.}
		\label{}
	\end{subfigure}
	\hfill
	\begin{subfigure}[b]{0.45\textwidth}
		\fbox{\includegraphics[width=\textwidth]{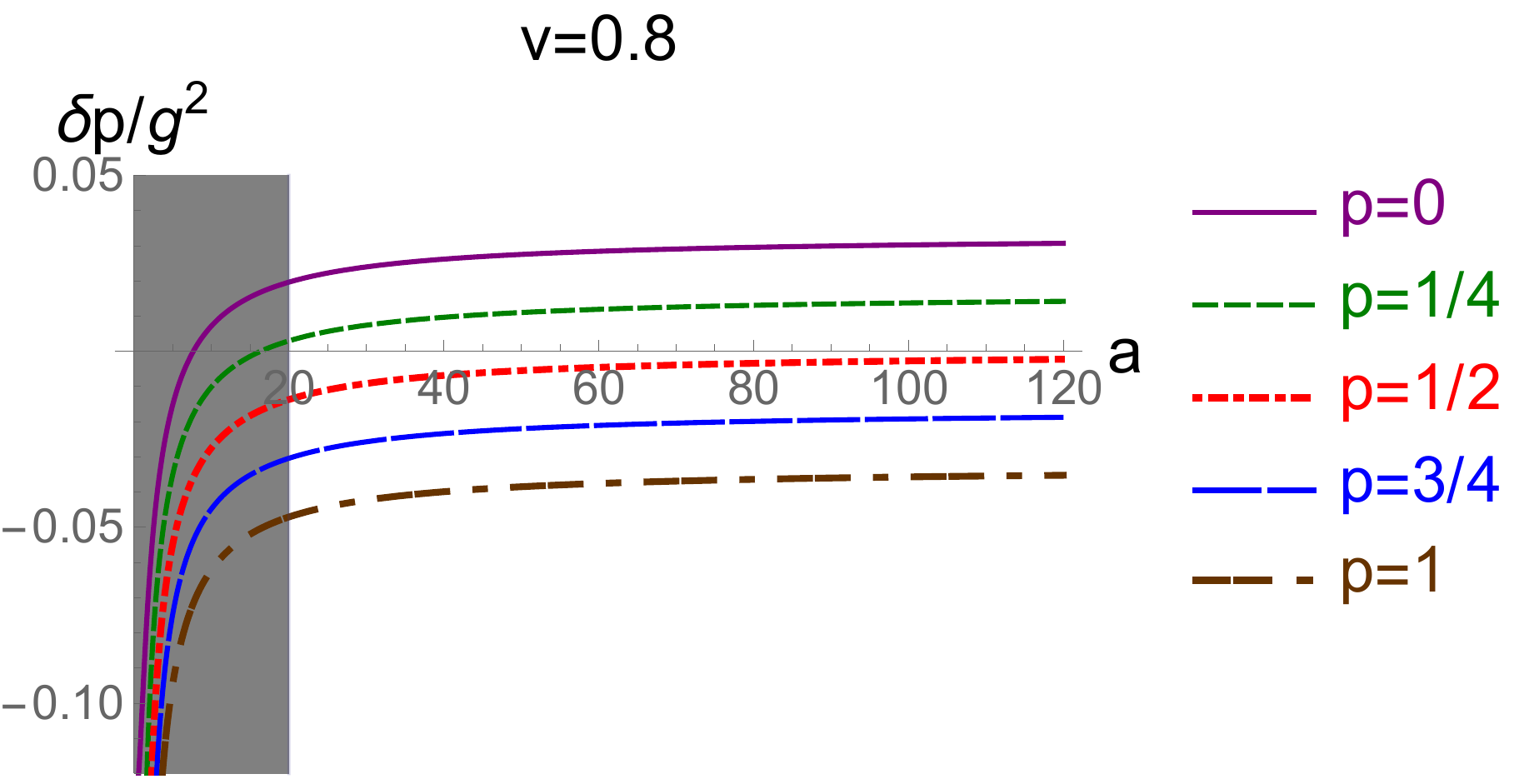}}
		\caption{Behavior for fixed velocity.}
		\label{}
	\end{subfigure}
	\caption{Behavior of the excited state population correction $\delta p$ as function of the reduced 
		acceleration $a=\alpha/\omega$.}
	\label{versus a}
\end{figure}

We will now numerically analyze the behavior of $\delta p$ given in Eq. (\ref{delta}) 
for different regimes of the Unruh machine. The behavior of $\delta p$ can be provided 
as a function of the reduced acceleration $a=\alpha/\omega$, for different values 
of initial probabilities $p$ and velocities $v$. The results are illustrated in Fig. (\ref{versus a}). 
Note that the corrections $\delta p$ always increase with the acceleration, which plays here 
the role of the quantum vacuum temperature. In particular, these corrections asymptotically tend to a 
velocity-dependent constant value for sufficiently high accelerations. Moreover, notice 
that perturbation theory requires that $| \delta p | \ll p$. This condition may be violated 
for small $a$, which indicates a breakdown of the perturbative approach. Indeed, as given 
by Eq.~(\ref{tva}), the interaction time $\tau$ tends to increase as we decrease $a$ for a fixed 
velocity $v$. Since the integral in Eq.~(\ref{integral}) implies that 
$| \delta p |  \propto g^2 \omega\tau p$, we have that a large $ | \delta p | $ may be yielded in the 
region of small $a$. This regime is forbidden by perturbation theory and is roughly indicated 
by the hatched gray areas in Fig. (\ref{versus a}). In fact, by imposing $| \delta p | \ll p$, for $p>0$, 
we then obtain the condition  $g^2 \omega\tau \ll 1$, which leads to
\begin{equation}
a \gg g^2\arctanh(v).
\label{acondition}
\end{equation}
Let us discuss in more details the special cases for initial probabilities $p=0$, $0<p<1/2$ and 
$1/2<p<1$, which are exhibited in Figs. (\ref{versus a}a, \ref{versus a}b, \ref{versus a}c), respectively. 
For the case $p=0$, we have that the excited state is initially empty and the qubit begins 
the interaction with the vacuum in its ground state. The breakdown of the perturbative approach 
appears here in the negative corrections $\delta p$ for small $a$ and this is represented by the gray area.
This area, between $a=0$ and $a=20$, encloses one order of magnitude greater that $g^2\arctanh(v)$, for $g=1$ and $v=0.99$,
as dictated by Eq. (\ref{acondition}). 
For larger values of $a$, the correction $\delta p$, 
gets positive and asymptotically increases for larger velocities $v$. 
This result is consistent with the Unruh effect since, if the system is prepared at 
the ground state, then the thermal fluctuations created by the Unruh effect would induce an 
increase in the excited state population. In addition, if $v$ is increased, this means a longer 
interaction time $\tau$ for a fixed acceleration $a$, which is expected to yield an asymptotically 
larger correction $\delta p$.   
 
For $p=1/4$, as shown in Fig. (\ref{versus a}b), we again see for high accelerations an asymptotically constant value of 
$\delta p$ that increases as a function $v$, as expected by the Unruh effect. For small acceleration 
values, we also observe the breakdown of the perturbative approach through the crossings of the curves 
for distinct values of $v$. The crossings would mean that different interaction times $\tau$ would 
imply the same excitation $\delta p$, which is a clearly a limitation of the perturbation theory. 
Moreover, for the point of vanishing acceleration, $a=0$, we see that a negative divergence appears. 
The origin of this behavior can be traced back to the the integral in Eq. (\ref{integral}), which is 
proportional to the interaction time interval. Since  $\tau\sim1/a$, then a divergence comes out for small $a$. 
This occurs because we would need an infinite time interval to change 
the velocity from $v$ to $-v$, by moving the qubit with zero acceleration. 

The case $p=3/4$ is presented at Fig. (\ref{versus a}c). We observe that, for this population of the 
excited state, the vacuum fluctuations would always cause a decrease of the original highly populated 
excited state. Then, spontaneous and stimulated decay in the qubit energy state would overcome the 
excitations caused by the Unruh effect. As the acceleration $a$ becomes larger, we have that the corrections 
$\delta p $ increase, but keeping always negative as a function of $a$.  

In Fig. (\ref{versus a}d), we show the behavior of $\delta p$ as a function of $a$ for different initial
populations $p$ of the excited state for a fixed velocity. This plot summarizes the previously discussed 
behaviors for both $p<1/2$ and $p>1/2$. For the special case $p=1/2$, any value of accelerations 
the Unruh thermal radiation produces a decay in the qubit energy state. However, in the limit of 
$a \rightarrow \infty$, the corrections are vanishing, i.e. $\delta p=0$. Then, in this case, high 
accelerations produce a balance on the original equally populated states and no net effect can 
be detected. In conclusion, it follows that, in order to gain heat from the quantum vacuum, we would require  
an initial state with $ 0 \leq p < 1/2 $. Moreover, perturbation theory requires (\ref{acondition}) as 
a validation approach.
\begin{figure}[hb]
	\begin{subfigure}[b]{0.48\textwidth}
		\fbox{\includegraphics[width=\textwidth]{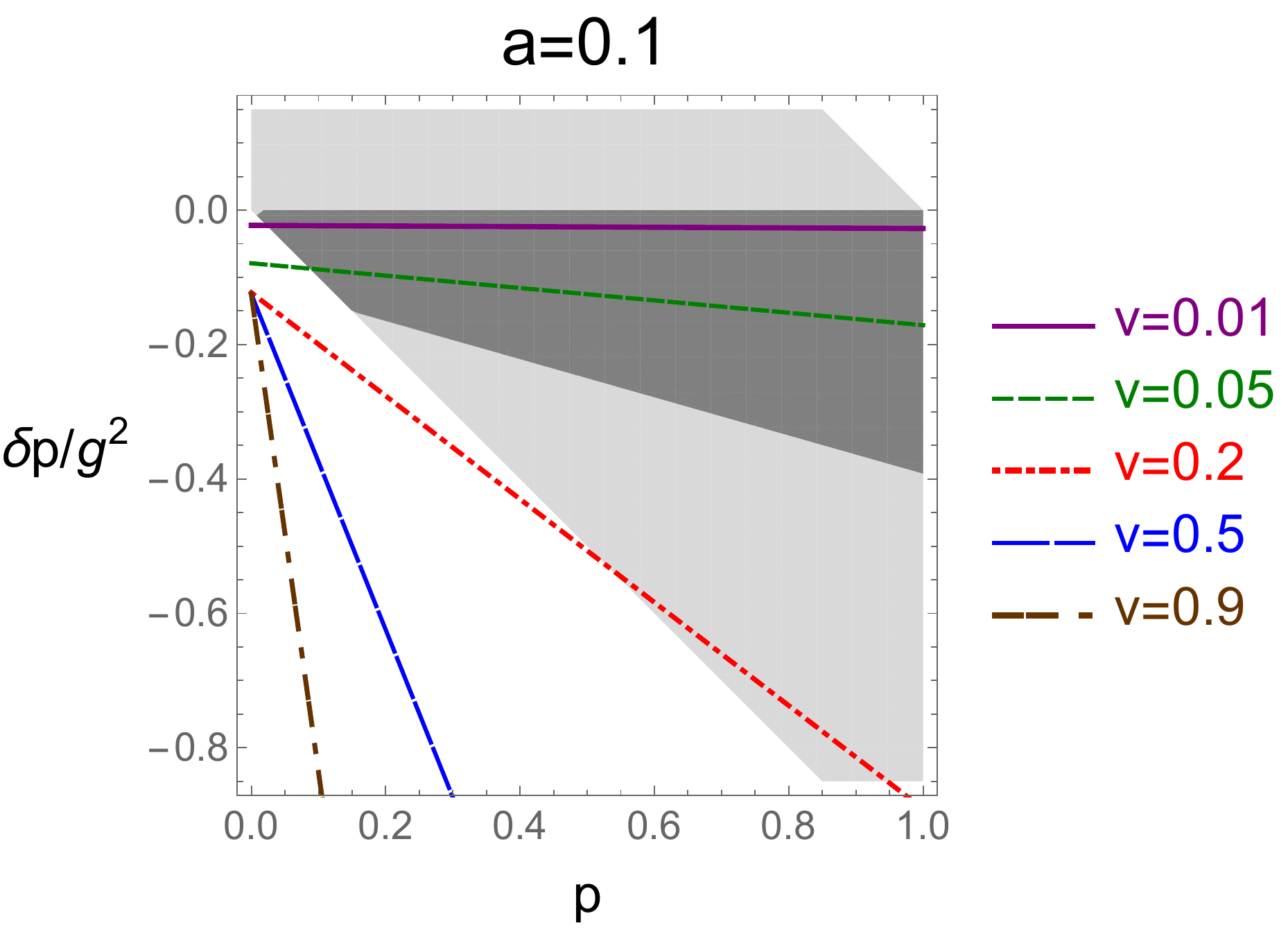}}
		\caption{Small accelerations.}
		\label{fig:f1}
	\end{subfigure}
	\hfill
	\begin{subfigure}[b]{0.49\textwidth}
		\fbox{\includegraphics[width=\textwidth]{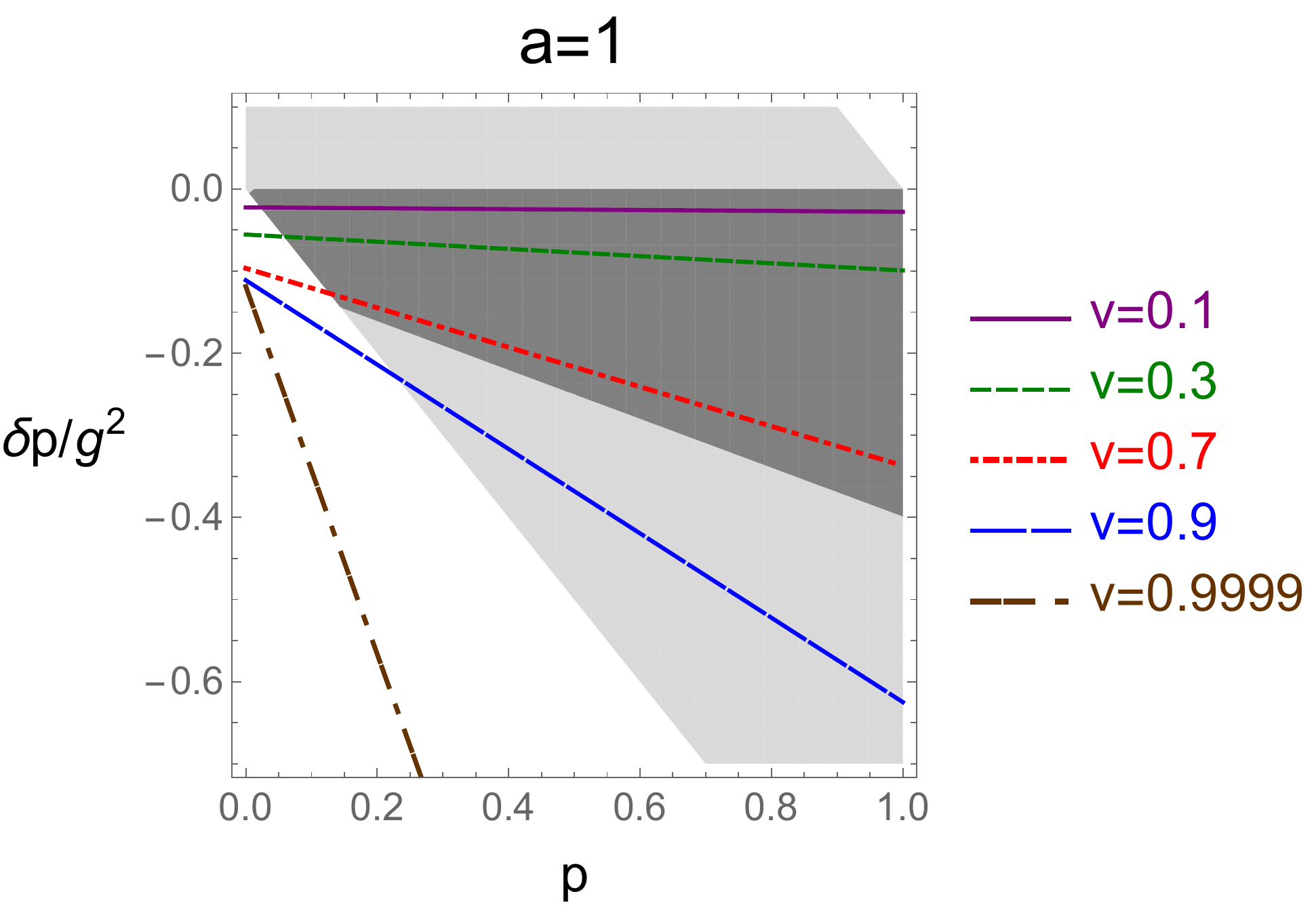}}
		\caption{Critical point $\mathrm{a}=1$, ($\alpha=\omega$).}
		\label{fig:f2}
	\end{subfigure}
	\hfill
	\begin{subfigure}[b]{0.48\textwidth}
		\fbox{\includegraphics[width=\textwidth]{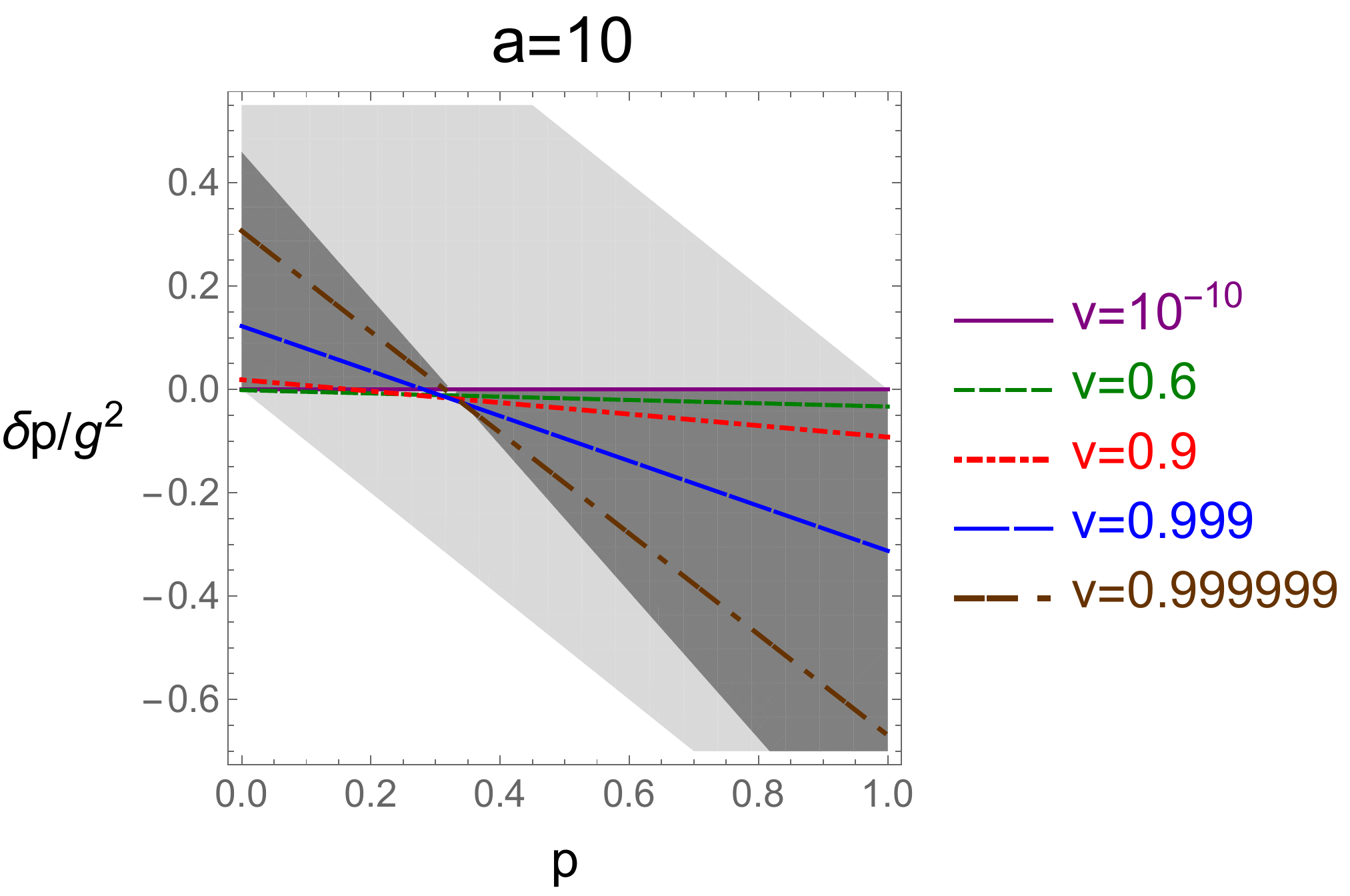}}
		\caption{High accelerations.}
		\label{fig:f2}
	\end{subfigure}
	\hfill
	\begin{subfigure}[b]{0.485\textwidth}
		\fbox{\includegraphics[width=\textwidth]{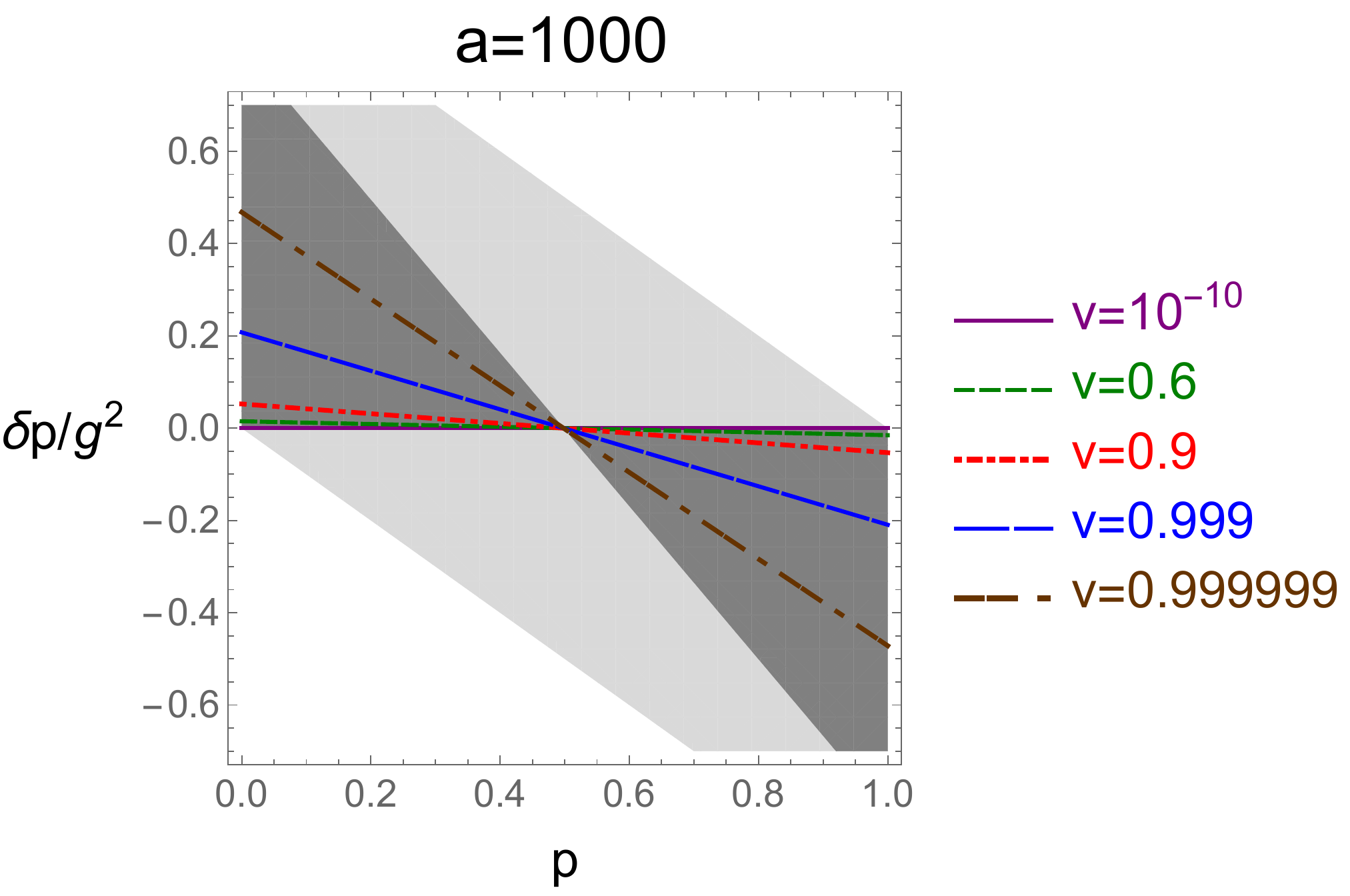}}
		\caption{Ultrahigh accelerations.}
		\label{fig:f2}
	\end{subfigure}
	\caption{Behavior of the corrections of the excited state population $\delta p$, in units of $g^2$, 
	as function of its initial population, $p$, for different accelerations and velocities. 
	The lighter gray area indicates corrections that obey $0<p+\delta p<1$, where $g=1$ has been considered.
	The darker gray area indicates points where the perturbative regime is satisfied, ie. $|\delta p|\ll p$.
	}
	\label{versus p}
\end{figure}

We now illustrate  in Fig. (\ref{versus p}) the behavior of $\delta p$ as a function of the initial population $p$ 
of the excited state. As it can be seen, the correction 
$\delta p$ is linear with $p$, as expected from Eq. (\ref{delta}). In addition, this linear response is such that 
$\delta p$ decreases as the initial population $p$ increases. This result is consistent with the Unruh effect 
over the qubit. For $p\approx0$, we have an initially lowly populated excited state, 
so that one expects that the Unruh effect yields an increase of its population, i.e. $\delta p>0$. On the other hand, 
for  $p\approx1$, the vacuum fluctuations are expected to yield a decay, i.e. $\delta p<0$. Therefore, as $p$ goes 
from $0$ to $1$, one expects the corrections $\delta p$ to decrease and invert its signal, as shown in Fig. (\ref{versus p}).

In all the plots of Fig. (\ref{versus p}), we also show a lighter gray area, which represents the allowed
values for the corrections $\delta p$ in order to get final positive probabilities less than one.
Then, this lighter gray area ensures $0<p+\delta p<1$, for illustration in this area it has been adopted $g=1$.
In order to evidence the limits of the perturbative approach we see that there exist values of $\delta p$
that are not in the lighter gray area. However, by using a lower value of the coupling constant as, for example, $g=0.1$, 
we get that the lighter gray area expands, making better the perturbative results.
Then, the lighter gray area is a necessary condition for the validity 
of perturbation theory. 
In Fig. (\ref{versus p}) the darker gray area indicates a stronger constraint over perturbation theory, provided 
by Eq.~(\ref{acondition}). This condition gives us the value of the maximum allowed velocity $v_{max}$ given a 
value for the acceleration $a$. From Eq.~(\ref{acondition}), we adopt $v_{max}=\tanh(a/g^2)$.
In Fig. (\ref{versus p}a), we have analyzed the case for a small reduced acceleration. 
In this case, we find that only very small values of the velocity remain in the domain of validity. As we increase 
acceleration, Fig. (\ref{versus p}b), we note that perturbation theory remains valid for a wider range of velocities 
and only very high velocities $v\sim1$ are capable to break down the perturbative results. The cases of high 
and ultrahigh accelerations, shown in Figs. (\ref{versus p}c) and (\ref{versus p}d), respectively, are rather robust 
against violations of the perturbative regime. 

\section{Efficiency and cyclicity of the Unruh heat engine}

\begin{figure}[t]
\centering
\includegraphics[scale=0.49]{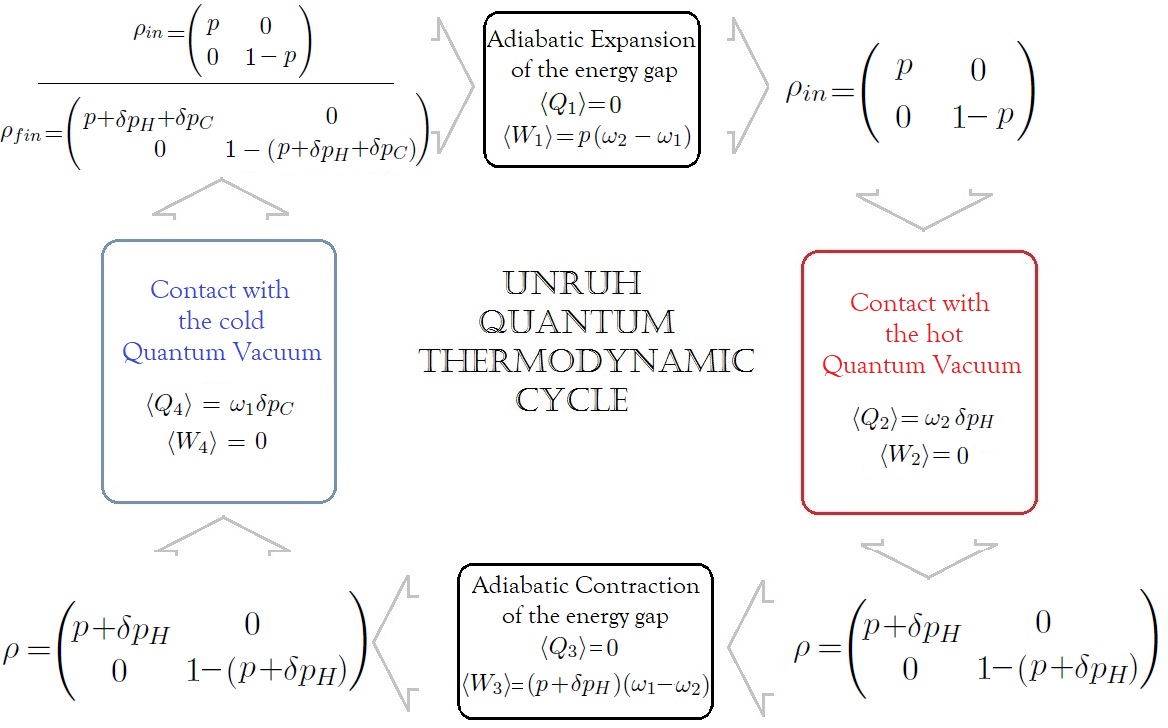}
\caption{Schematic design of the thermodynamic cycle for the Unruh quantum heat engine. 
The figure illustrates the different states of the qubit, the absorbed heat, and the produced work 
at each stage of the cycle.}
\label{totalcycle}
\end{figure}
We can summarize the Unruh heat engine cycle as shown in Fig. (\ref{totalcycle}).
There, we explicitly exhibit the absorbed heat and produced work  at each part of the cycle, 
as well as the changes in the qubit density operator. As we mentioned before, the first 
contact of the qubit with the quantum vacuum occurs as it moves at a high acceleration $\alpha_H$. 
In this part of the cycle the qubit should absorb heat $\langle Q_2\rangle>0$ from the quantum 
vacuum and increase the population of the excited state by $\delta p_H> 0$. After this first contact, 
the qubit undergoes a contraction of the energy gap. Then, it is put in contact with the quantum 
vacuum again. During the second contact of the qubit with the quantum vacuum, it should transfer 
heat $\langle Q_4\rangle<0$ and changes its excited state population by $\delta p_C<0$. 
Therefore, in order for the cycle to yield positive net work,
we need to assure both that the vacuum behaves like the hot thermal reservoir in the first contact 
with the qubit and that it acts as a cold thermal reservoir in the second contact with the qubit. 
We can explicitly analyze the regimes for these conditions to be obeyed. 
As we previously mentioned, to obtain a closed cycle, we require that the qubit comes back to its 
original state as the machine returns to its first stage, which means 
\beq
\delta p_H+\delta p_C=0.
\label{cond}
\eeq
Since we expect that $\delta p_H>0$, then we should also have $\delta p_C<0$. 
The periodic condition given by Eq. (\ref{cond}) is equivalent to require that the 
first law of thermodynamics remains valid on average for our process, as we will see. 
Consequently, this
leads us to the total mean quantum heat absorbed by 
the qubit from the vacuum
\bea
\langle Q\rangle&=&\langle Q_2\rangle+\langle Q_4\rangle,\nn\\
&=&\omega_2\delta p_H+\omega_1\delta p_C,\nn\\
&=&(\omega_2-\omega_1)\delta p_H.
\eea
On the other hand, work is realized only during the adiabatic expansion or 
contraction of its energy gap. Hence, the total work produced over the qubit is
\bea
\langle W\rangle&=&\langle W_1\rangle+\langle W_3\rangle,\nn\\
&=&(\omega_2-\omega_1)p+(\omega_1-\omega_2)(p+\delta p_H),\nn\\
&=&(\omega_1-\omega_2)\delta p_H.
\eea
Then, as one can see, the net work produced over the qubit and the total heat absorbed from
the vacuum satisfies $\langle Q\rangle+\langle W\rangle=0$, as is expected from the first
law of thermodynamics. Since $\langle W\rangle$ is the work performed over the 
qubit by external forces, then the work performed by the qubit over the exterior is given by 
$\langle W\rangle_{ext}=-\langle W\rangle$.
Hence, we define the efficiency $\eta$ as
\beq
\eta=\frac{\langle W\rangle_{ext}}{\langle Q_2\rangle},
\eeq
which yields
\beq
\eta=1-\frac{\omega_1}{\omega_2}.
\eeq
The efficiency depends only on the ratio of the minimum and maximum energy gaps 
and is independent of the accelerations (temperatures) we have considered.
Moreover, this is the same efficiency found by using a classical thermal bath~\cite{Kieu04},
which suggestes a kind of universal bound independently of the nature of the thermal reservoir.
\begin{figure}[t]
	\centering
		\includegraphics[scale=0.5]{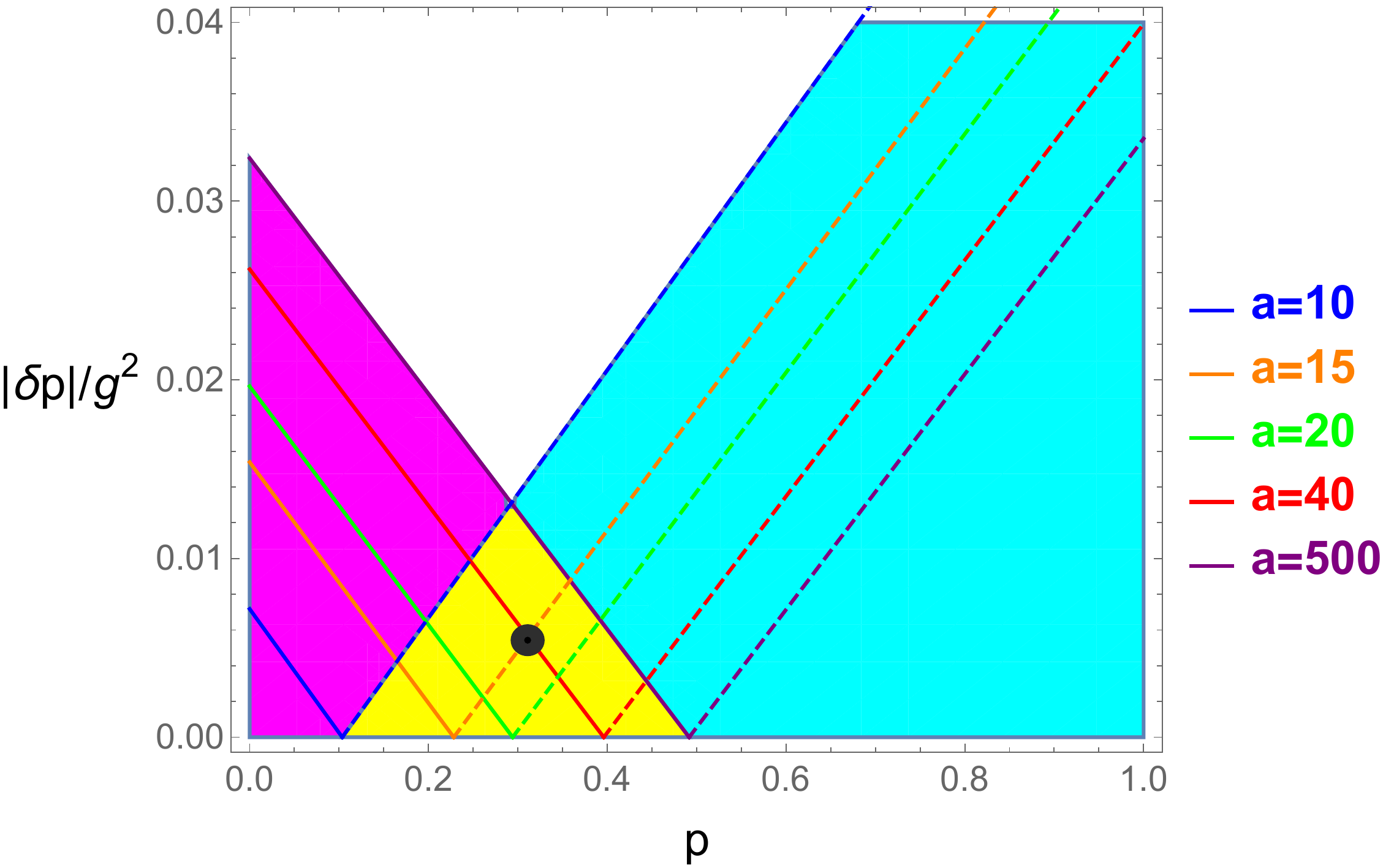}
			\caption{This plot shows the behavior of the absolute value of the correction $|\delta p|$ 
			as function of $p$, for various  reduced accelerations $a=\alpha/\omega$ and velocity $v=0.8$. Continuous lines 
			contain points representing positive corrections, with the vacuum behaving as a hot thermal bath. 
			Dashed lines show negative corrections, with the vacuum acting as a cold thermal bath.
			For the magenta region, the vacuum only acts as a hot reservoir. For the cyan region, 
			the vacuum only acts as a cold reservoir. In the yellow region, we show that the cycle 
			can be closed, since the vacuum acts either as hot or cold reservoir for the same initial value 
			of $p$, with hot or cold depending on the different reduced accelerations associated with the trajectory. 
			The black point intersection defines a triplet ($a_H$, $a_C$, p) for which we can define a closed cycle.}
			\label{domains}
\end{figure}

Let us now, proceed to analyze the kinematic constraints for the cyclic condition
of the qubit state, Eq. (\ref{cond}). By defining the hot $a_H$ and cold $a_C$ reduced accelerations for the qubit 
in its two contacts with the vacuum  as $a_H=\alpha_H/\omega_2$ and $a_C=\alpha_C/\omega_1$ and using the 
general result Eq. (\ref{delta}), then the cyclicity condition given by Eq. (\ref{cond}) can be written as 
\beq
\delta  p(a_H,p,v)+\delta  p(a_C,p,v)=0.
\label{eqpc}
\eeq
We also have to ensure that, during the first contact with the vacuum, the qubit has a larger energy gap 
$\omega_2$ and that the qubit absorb heat from the vacuum fluctuations, so that $\delta p(a_H,p,v)>0$. 
This, together with the cyclic condition Eq. (\ref{eqpc}),  would imply that during the second contact with the vacuum, were the qubit
has a lower energy gap $\omega_1$, the vacuum acts as a thermal sink, $\delta  p(a_C,p,v)<0$.
Then, for the same values of $p$ and $v$, but different values of reduced accelerations $a_H$ and $a_C$, we must ensure that
the vacuum acts like a hot or cold thermal bath, respectively. 
In Fig. (\ref{domains}) we show the behavior of the absolute value of the correction $|\delta p|$
as a function of the initial probability $p$, for a velocity $v=0.8$ and different reduced accelerations.
In solid lines, we show the points where the qubit absorbs heat from the vacuum ($\delta p>0$) and, in dashed lines,
the points where the qubit emits heat to the vacuum ($\delta p<0$). The yellow region represents the possible values of 
$a_H$, $a_C$ and $p$, where we can find an intersection between solid lines and dashed lines.
Therefore, these regions show points corresponding to pairs of accelerations for which, by choosing the initial probability of the 
excited state as $p$, we will have a closed thermodynamic cycle and extract work from the vacuum. As an illustration, the intersection 
of the red solid line ($a_H=40$) and the dashed orange line ($a_C=15$) occurs at the value $p=0.293$. This triplet satisfies 
Eq. (\ref{eqpc}), with $v=0.8$, allowing for a closed cycle in the Unruh heat engine.\\

\begin{figure}[t]
  \begin{subfigure}[b]{0.47\textwidth}
    \includegraphics[width=\textwidth]{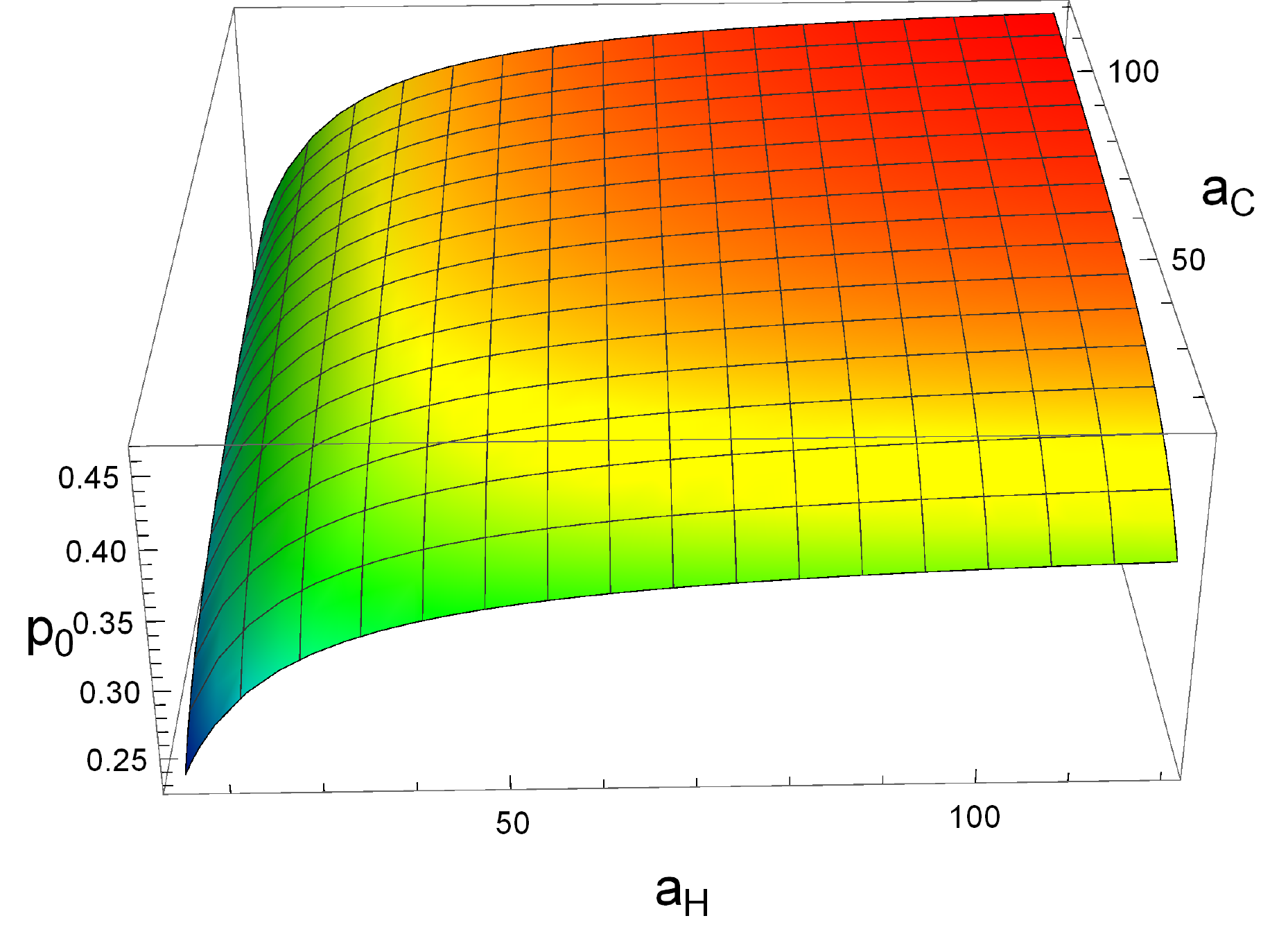}
    \caption{Critical value of the initial probability $p_0$ as function of $a_H$ and $a_C$, for $v=0.8$, 
    such that we have a closed Unruh cycle.}
    \label{}
  \end{subfigure}
	 \hfill
  \begin{subfigure}[b]{0.46\textwidth}
    \includegraphics[width=\textwidth]{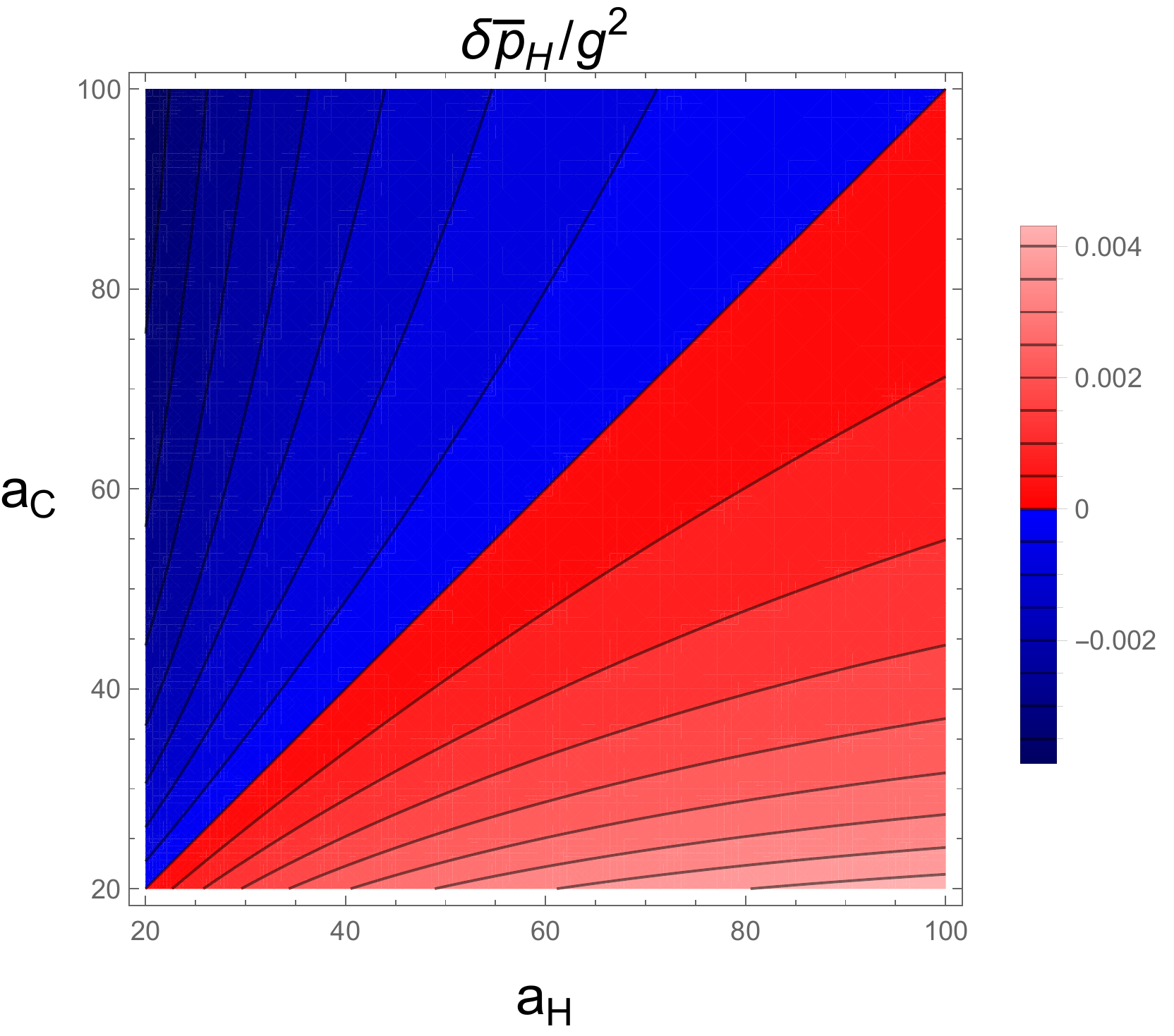}
    \caption{Correction $\delta \overline{p}_H$ for different accelerations $a_H$ and $a_C$. 
		We get positive heat from the quantum vacuum $\delta \overline{p}_H>0$, only for $a_H>a_C$.}
    \label{}
  \end{subfigure}
  \caption{}
	\label{cyclic}
\end{figure}
\noindent
In a more general analysis, Eq. (\ref{eqpc}) can be interpreted as an equation for finding the initial population of the 
excited state $p$, in such a way that the vacuum acts as a hot reservoir for the qubit with reduced acceleration $a_H$, 
while at the same value of $p$, the vacuum acts as a cold reservoir for the qubit with reduced acceleation $a_C$. 
In this manner, we define the critical probability $p_0$ that solves Eq. (\ref{eqpc}) as the value for our initial population 
such that the cycle is closed and the qubit returns to its initial state. Then, this probability, which represents intersections 
of solid and dashed lines in Fig.~\ref{domains}, depends on the values of the hot and cold 
reduced accelerations (vacuum temperatures), reading
\beq
p_0=p_0(a_H,a_C,v).
\eeq
By using Eq. (\ref{delta}) into Eq.~(\ref{eqpc}) and defining the function
\beq
{\cal P}(a_H,a_C,v)=\frac{2a_H\cdot a_C}{(a_H+a_C)\arctanh(v)}\left(J\left(-\frac{1}{a_H},2\arctanh(v)\right)+J\left(-\frac{1}{a_C},2\arctanh(v)\right)\right),
\eeq
we can shown that the solution for the initial critical probability $p_0$ is
\beq
p_0=\frac{{\cal P}}{1+2{\cal P}}.
\label{pc}
\eeq
The behavior of the solution given by Eq. (\ref{pc}) for a different pair of reduced accelerations is shown in Fig. (\ref{cyclic}a).
There, we can see that $0<p_0<1/2$, as expected from previous discussions.
Now, using this solution we can define the correction $\delta \overline{p}_H$ such that
\beq
\delta \overline{p}_H=\delta \overline{p}_H(a_H,a_C,v)=\delta p(a_H,p_0,v).
\eeq
The behavior of this function is shown in Fig. (\ref{cyclic}b). 
There, we observe that heat is absorbed from 
the hot quantum vacuum, $\delta \overline{p}_H>0$, and the cycle behaves as a thermal machine 
instead of a refrigerator, only when $a_H>a_C$, which implies in the condition
\beq
\alpha_H>\alpha_C\left(\frac{\omega_2}{\omega_1}\right).
\label{des}
\eeq
This relation is analog to the result find in \cite{Kieu04}, however in our case the role of temperatures 
is interpreted by the accelerations of the qubit when interacts with the vacuum fluctuations.
This is in accordance with the thermal nature of the quantum vacuum predicted by Unruh.
This condition, Eq. (\ref{des}), is stronger that the classical relation $\alpha_H>\alpha_C$, 
and establish which gradients of accelerations (temperatures) could be used in order to run a quantum heat engine.
Note also that 
we have find the initial critical probability, given by Eq. (\ref{pc}),
in order to run the quantum thermal machine. 
Finally, by defining the change due to the cold reservoir as $\delta \overline{p}_C=\delta p(a_C,p_0,v)$,
we get that  $\delta \overline{p}_C = - \delta \overline{p}_H$, so in the domain defined by  Eq. (\ref{des})
we obtain that the vacuum acts like a cold reservoir when the qubit is prepare 
to have initial critical probability $p_0$ and has a small reduced acceleration $a_C$. 
All this ensure us that the cycle behaves as a heat engine instead of a refrigerator.

\begin{figure}[t]
	\centering
		\includegraphics[scale=0.4]{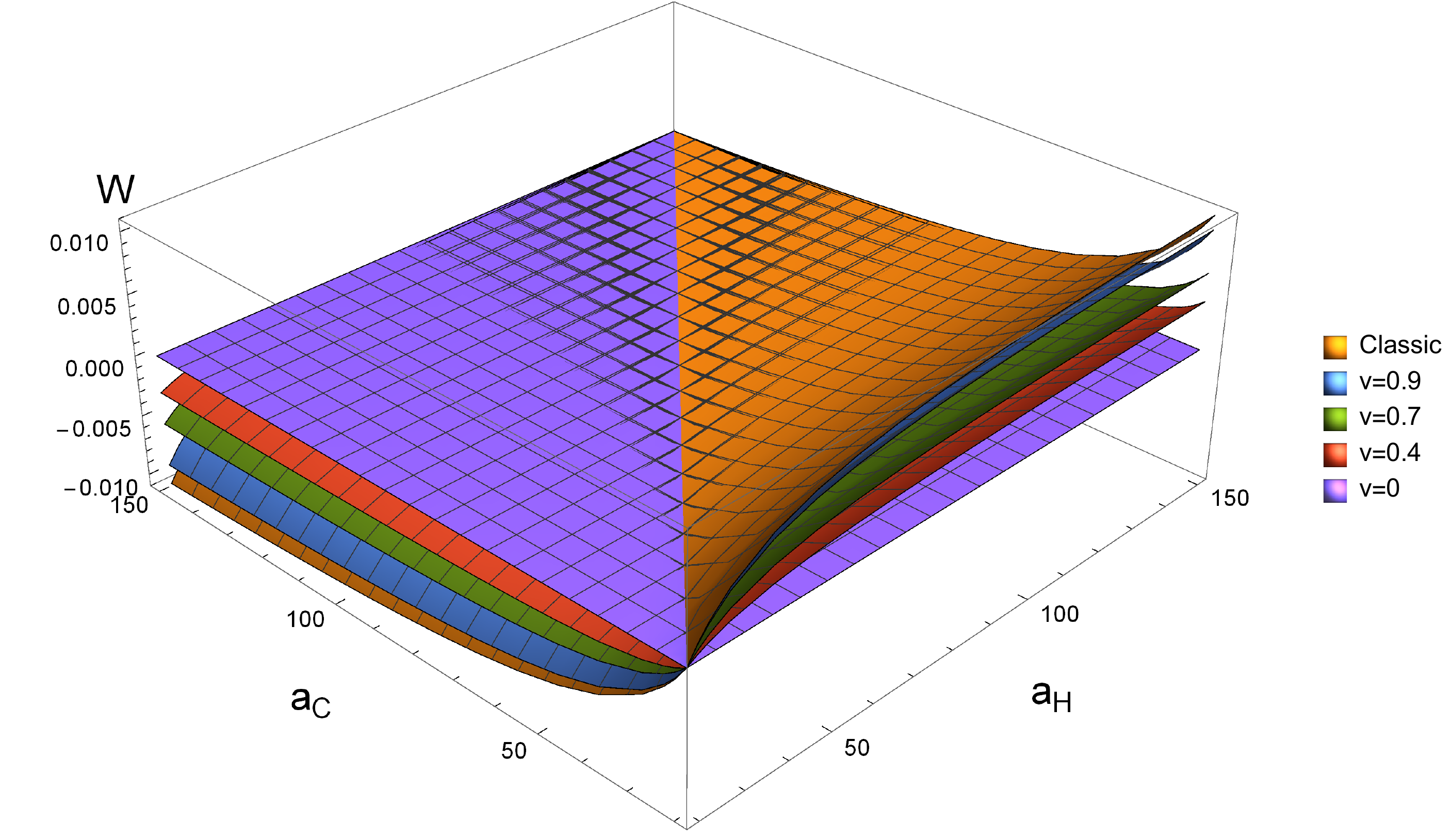}
			\caption{Comparison between the work performed by the Unruh heat engine with different qubit velocities and the usual
			quantum heat engine with a classical thermal reservoir. Here we have used $\omega_2-\omega_1=1.$}
			\label{classicbath}
\end{figure}

In order to conclude our analysis, we will realize a comparison between the works performed by 
Unruh quantum Otto engine and the usual quantum heat engine relying on a classical thermal bath. 
For a classical thermal bath, we have that the work provided by the machine is given by 
$W_{cl}=(\omega_2-\omega_1)\delta p_{cl}$, where
\beq
\delta p_{cl}=\frac{1}{1+e^{\beta_H\omega_2}}-\frac{1}{1+e^{\beta_C\omega_1}},
\eeq
is the change in the population of the excited state when the qubit pass through the hot thermal bath. 
Here we suppose that the qubit begins this contact with an equilibrium distribution, which has a temperature 
given by the cold reservoir and, after this contact, the qubit state has an equilibrium distribution with temperature given by the hot reservoir.
We compare our result $W=(\omega_2-\omega_1)\delta \overline{p}_{H}$ with the classical result when the temperatures $T_{H,C}=1/\beta_{H,C}$
coincide with the accelerations $\alpha_{H,C}$ we used here. Then, in this case, we can rewrite the classical correction in terms of the reduced accelerations
\beq
\delta p_{cl}(a_H,a_C)=\frac{1}{1+e^{1/a_H}}-\frac{1}{1+e^{1/a_C}}.
\eeq
Since, our correction $\delta\overline{p}_{H}(a_H,a_C,v)$ also depends on the velocity of the qubit in the kinematic cycle, 
we can compare this correction for different values of the qubit velocity versus the classical thermal bath correction. 
This is shown in Fig. (\ref{classicbath}). There we can see that, as we increase the qubit velocity $v$ 
in the quantum Unruh heat engine, the results approach the expected result of a classical thermal bath.
This can be understood in terms of the interaction time with the vacuum. Since we are dealing with a	 finite trajectory
in space-time, the interaction time of the qubit with the vacuum fluctuations is finite and given by ${\cal T}=2\arctanh(v)/\alpha$.
So, as we increase the qubit velocity, the interaction time  with the vacuum fluctuations also increases and we approach a
classical thermal equilibrium. It is worth noting here that perturbation theory constrains us to small values of ${\cal T}$,
so very small values of $\alpha$ as well as ultrahigh velocities $v\approx 1$ must be avoided.

\section{Conclusions}
We have introduced a relativistic quantum thermal machine based on the Unruh effect. This is achieved 
by taking the quantum vaccuum as a thermal reservoir. By using quasi-static processes and perturbation 
theory, we are then able to establish the conditions over the initial excitation probability $p_0$ and qubit 
accelerations $a_H$ and $a_C$  for ensuring closed cycles, both from the thermodynamic and kinematic 
point of view. More specifically, we determine sets of triplets $(p_0,a_H,a_C)$ such that a closed cycle emerges. 
Moreover, we have also shown that the classical efficiency 
of the Otto cycle persists in the relativistic regime, suggesting a universal bound as long as quasi-static 
processes are adopted. On the other hand, work is nontrvially affected by kinematics, being  
dependent on the velocities and accelerations of the qubit throughout its time evolution. This analysis has 
been analytically provided via perturbation theory, with its limitations delineated. 

Even though realizations of the Unruh effect are still challenging, the thermal machine proposed 
here opens the possibility of posing relativistic quantum themodynamics in an experimental setting 
for quantum motors and refrigerators as long as alternative approaches~\cite{Felicetti:15,Laguna:17,Wang:14,Ahluwalia:16,Cozzella:17,Crispino:08} 
for probing the Unruh effect are well-succeeded. As further developments, it remains the extension of 
our approach to the non-equilibrium regime in open systems, e.g., with the entropy production taken 
into account~\cite{Deffner:11}. Moreover, it is still promising the investigation of the role of correlations, 
such as entanglement, for the machine efficiency. These topics are left for future research.

\section*{Acknowledgments}

E.A. would like to thank Guillermo Due\~nas for many useful discussions while this work has been prepared
and to FAPERJ for financial support.
T.R.O. would like to thank CNPq-Brazil for financial support.
M.S.S. acknowledges support from CNPq-Brazil 
(No.303070/2016-1) and FAPERJ (No 203036/2016). The authors also acknowledge the Brazilian National Institute 
for Science and Technology of Quantum Information (INCT-IQ).


\appendix
\numberwithin{equation}{section}
\begin{appendices}

\section{Dynamics of the qubit-vacuum interaction}
\label{ap:dynamics}

During the interaction of the qubit with the quantum scalar field
we have that the Hamiltonian of the total qubit-field system 
is given by
\beq
\mathbb{H}=\mathbb{H}_0+\mathbb{H}_{int},
\eeq
the free Hamiltonian $\mathbb{H}_0$ is the part that does not consider interaction between 
the qubit and the field 
\beq
\mathbb{H}_0={\cal H}+{\cal H}^{field},
\eeq
where
${\cal H}^{field}=\int d^3x (1/2)\{(\partial_t\varphi)^2+(\nabla\varphi)^2\}$
is the Klein-Gordon Hamiltonian associated to a free massless scalar field $\varphi$
and ${\cal H}$ is the the qubit free Hamiltonian given by 
${\cal H}=\omega\,|e\rangle \langle e|$. The qubit-field interaction Hamiltonian
is given by 
\beq
\mathbb{H}_{int}= g\,m\varphi(\chi(\tau)),
\eeq
where $g$ is a small coupling constant of the interaction,
$m$ is the qubit monopole operator given by
\beq
m=|e\rangle\langle g|+|g\rangle\langle e|,
\eeq
and $\varphi(\chi(\tau))$ is the scalar field evaluated on the spacetime point $\chi(\tau)=(t,x)$,
where the qubit is located. These coordinates are given by Eq. (\ref{tx}),
with the qubits moving with constant acceleration $\alpha$.
Let us denote $\varrho$ the total density operator of the qubit-field system. 
Then this operator satisfies the quantum Liouville's equation
\beq
i\frac{d\varrho(t)}{dt}=[\mathbb{H},\varrho].
\label{sch}
\eeq
The qubit is prepared in the mixed state $\rho_{in}$ and the scalar field starts  
at the pure vacuum state $|0\rangle$. The initial density operator of the total 
system is taken as the tensor product  
$\varrho_{in}=\rho_{in}\otimes|0\rangle\langle0|$.
In order to simplify the dynamics, we move to the interaction 
picture. Then, we define the total density operator in the
interaction picture $\varrho_I(t)$ as a pull-back over $\varrho(t)$, yielding
\beq
\varrho_I(t)=\mathbb{U}_0^{-1}(t,t_0)\varrho(t)\mathbb{U}_0(t,t_0),
\eeq
where $\mathbb{U}_0(t,t_0)$ is the total evolution operator when there is no 
interaction between the qubit and the field. Therefore, this operator satisfies
\beq
i\frac{\partial\mathbb{U}_0(t,t_0)}{\partial t}=\mathbb{H}_0\mathbb{U}_0(t,t_0).
\eeq
The dynamics of the density operator in the interaction picture then reads
\beq
i\frac{d\varrho_I(t)}{dt}=[\mathbb{H}_{int}^I,\varrho_I(t)],
\label{schin}
\eeq
where $\mathbb{H}_{int}^I$ is the interaction Hamiltonian in the interaction picture.
As before, the initial condition is given by 
$\varrho_I(t_0)=\varrho(t_0)=\varrho_{in}=\rho_{in}\otimes|0\rangle\langle0|$.
Hence, the solution to Eq. (\ref{schin}) is given in terms of a Dyson series 
\bea
&&\!\!\!\!\!\!\!\!\!\!\varrho_I(t)=\varrho_{in}-i\int_{t_0}^td\tau\left[\mathbb{H}_{int}^I(\tau),\varrho_{in}\right]
-\int_{t_0}^td\tau\int_{t_0}^\tau d\tau'\left[\mathbb{H}_{int}^I(\tau),\left[\mathbb{H}_{int}^I(\tau'),\varrho_{in}\right]\right]+\ldots 
\eea
We use now the temporal order product operator $T$, which acts as
\bea
T\left\{\left[\mathbb{H}_{int}^I(\tau),\left[\mathbb{H}_{int}^I(\tau'),\varrho_{in}\right]\right]\right\}&=&
\theta(\Delta\tau)\left[\mathbb{H}_{int}^I(\tau),\left[\mathbb{H}_{int}^I(\tau'),\varrho_{in}\right]\right]\nn\\
&&+\theta(-\Delta\tau)\left[\mathbb{H}_{int}^I(\tau'),\left[\mathbb{H}_{int}^I(\tau),\varrho_{in}\right]\right],
\eea
with $\Delta\tau=\tau-\tau'$. Then, we can write $\varrho_I(t)$ as
\bea
\varrho_I(t)&=&\varrho_{in}-i\int_{t_0}^td\tau\left[\mathbb{H}_{int}^I(\tau),\varrho_{in}\right]\nn\\
&&-\frac{1}{2}\int_{t_0}^td\tau\int_{t_0}^t d\tau'T\left\{\left[\mathbb{H}_{int}^I(\tau),\left[\mathbb{H}_{int}^I(\tau'),\varrho_{in}\right]\right]\right\}+\ldots
\label{dyson}
\eea
We consider here that the contact with the vacuum occurs at general times 
$-\bar\tau<\tau<\bar\tau$, where  we have that $\bar\tau=\arctanh(v)/\alpha$. 
From Eq. (\ref{dyson}), we need to express 
the interaction Hamiltonian, $\mathbb{H}_{int}=g\,m\varphi(\chi(\tau))$,  in the interaction picture.
Given an operator in the Schrodinger picture ${\cal O}_S$, 
its version in the interaction
picture is
\beq
{\cal O}_I(t)=\mathbb{U}_0^{-1}(t, t_0){\cal O}_S\mathbb{U}_0(t, t_0).
\eeq
Therefore the interaction Hamiltonian in the interaction picture is given by
\beq
\mathbb{H}_{int}^I(\tau)=g\,\mathbb{U}_0^{-1}(\tau, t_0)\,\,m\varphi(\chi(\tau))\,\,\mathbb{U}_0(\tau, t_0),
\eeq
where we have defined 
\beq
\mathbb{U}_0(\tau, t_0)=e^{-i\mathbb{H}_0(\tau-t_0)}.
\eeq
Since $\mathbb{H}_0={\cal H}+{\cal H}^{field}$ is time independent and $\left[{\cal H},{\cal H}^{field}\right]=0$, 
it follows
\beq
\mathbb{U}_0(\tau, t_0)={\cal U}(\tau, t_0)\,{\cal U}^{field}(\tau, t_0),
\eeq
where ${\cal U}(\tau, t_0)=e^{-i{\cal H}(\tau-t_0)}$ and ${\cal U}^{field}(\tau, t_0)=e^{-i{\cal H}^{field}(\tau-t_0)}$.
Therefore
\beq
\mathbb{H}_{int}^I(\tau)=g\,m_I(\tau)\varphi_I(\chi(\tau)),
\eeq
where we have defined 
\bea
m_I(\tau)&=&{\cal U}^{-1}(\tau, t_0)\,m\,\,{\cal U}(\tau, t_0),\nn\\
\varphi_I(\chi(\tau))&=&{{\cal U}^{field}}^{-1}(\tau, t_0)\,\varphi(\chi(\tau))\,{\cal U}^{field}(\tau, t_0).
\label{mI}
\eea
Since operators in the interaction picture,
Eq. (\ref{mI}), evolve as free operators without taking into account the influence
between the field and the quibt, then 
the scalar field $\varphi_I(\chi(\tau))$ represents the free scalar field evaluated
at the position of the qubit. The vacuum expected values of products of this operator 
give us the free Green correlation functions of the field. For instance, the free Wightman
function of the free scalar field is given by
\beq
G^+(\chi,\chi 	')=\langle 0|\varphi_I(\chi)\varphi_I(\chi')|0\rangle.
\eeq
Let us now explicitly work the monopole operator at the interaction picture
$m_I(\tau)$. As the initial time during the contact with the vacuum is $t_0=-\bar\tau$ and the Hamiltonian
of the qubit ${\cal H}=\omega\,|e\rangle \langle e|$ is time independent,
then
\bea
{\cal U}_0(\tau,t_0)&=&{\cal U}_0(\tau,-\bar\tau),\nn\\
&=&e^{-i\omega_2(\tau+\bar\tau)\tilde\sigma},\nn\\
&=&\mathbb{I}+\left(e^{-i\omega_2(\tau+\bar\tau)}-1\right)\tilde\sigma,\nn\\
&=&\left(
\begin{array}{cc}
	e^{-i\omega_2(\tau+\bar\tau)} & 0 \\
	0 & 1
\end{array}
\right).
\label{u01q}
\eea
The monopole operator is given by $m=|e\rangle\langle g|+|g\rangle\langle e|$. Then its matrix representation is
\beq
m=\left(
\begin{array}{cc}
	0 & 1 \\
	1 & 0
\end{array}
\right).
\eeq
In the interaction picture $m_I(\tau)$ reads
\bea
m_I(\tau)&=&{\cal U}^{-1}(\tau,-\bar\tau)\,m\,\,{\cal U}(\tau,-\bar\tau),\nn\\
&=&\left(
\begin{array}{cc}
	0 & e^{i\omega_2(\tau+\bar\tau)} \\
	e^{-i\omega_2(\tau+\bar\tau)} & 0
\end{array}
\right).
\eea
The state of the qubit-field system at the instant $-\bar\tau<t<\bar\tau$ after the interaction
have begun is, up to second order in perturbation theory, given by
\bea
\varrho_I(t)&=&\varrho_{in}-ig\int_{-\bar\tau}^td\tau\left[m_I(\tau)\varphi_I(\chi(\tau)),\varrho_{in}\right]\nn\\
&&\!\!\!-\frac{g^2}{2}\int_{-\bar\tau}^td\tau\int_{-\bar\tau}^td\tau'\,
T\left\{\left[m_I(\tau)\varphi_I(\chi(\tau)),[m_I(\tau')\varphi_I(\chi(\tau')),\varrho_{in}]\right]\right\}.\nn\\
\eea
Since we are interested only in the qubit state independently of the field, we perform a partial trace
over the field degrees of freedom on the total state of the qubit-field system $\varrho_I(t)$, which yields
\beq
\rho_I(t)=\tr_{\mathrm{field}}\,\varrho_I(t).
\label{ro1q}
\eeq
By taking into account that $\varrho_{in}=\rho_{in}\otimes|0\rangle\langle 0|$, the partial traces produce
\bea
&&\tr_{\mathrm{field}}\,\varrho_{in}=\rho_{in},\nn\\
&&\tr_{\mathrm{field}}\left[m_I(\tau)\varphi_I(\chi(\tau)),\varrho_{in}\right]=0,
\label{tem1}
\eea
\bea
\tr_{\mathrm{field}}\left[m_I(\tau)\varphi_I(\chi(\tau)),[m_I(\tau')\varphi_I(\chi(\tau')),\varrho_{in}]\right]
&=&m_I(\tau)m_I(\tau')\rho_{in}G^+_{\alpha}(\tau,\tau')\nn\\
&+&\rho_{in}m_I(\tau')m_I(\tau)G^+_{\alpha}(\tau,\tau')\nn\\
&-&m_I(\tau)\rho_{in}m_I(\tau')G^+_{\alpha}(\tau,\tau')\nn\\
&-&m_I(\tau')\rho_{in}m_I(\tau)G^+_{\alpha}(\tau,\tau'),\nn\\
\label{tem2}
\eea
where we have evaluated the Wightman functions of the scalar field at different points of the qubit trajectory, 
denoting them as $G^+_{\alpha}(\tau,\tau')=G^+(\chi(\tau),\chi(\tau'))$.
In this notation we make explicit the fact that the Whightman function depends only on the difference of 
proper times and that the trajectory have a proper acceleration $\alpha$. 
In this way and by using the properties of the temporal ordered products, 
we obtain that the final state of the qubit in the interaction picture, Eq. (\ref{ro1q}), is given by
\beq
\rho_I(t)=\rho_{in}+g^2\int_{-\bar\tau}^td\tau\int_{-\bar\tau}^td\tau'\delta\rho(\tau,\tau')G^+_{\alpha}(\tau,\tau'),
\eeq
where we have defined
\beq
\delta\rho(\tau,\tau')=\delta\rho^{(a)}(\tau,\tau')+\delta\rho^{(b)}(\tau,\tau')+\delta\rho^{(b)}(\tau,\tau'),
\eeq
with each term that corrects the state of the qubit given by
\bea
\delta\rho^{(a)}(\tau,\tau')&=&m_I(\tau')\rho_{in}m_I(\tau),\nn\\
\delta\rho^{(b)}(\tau,\tau')&=&-\theta(\Delta\tau)m_I(\tau)m_I(\tau')\rho_{in},\nn\\
\delta\rho^{(c)}(\tau,\tau')&=&-\theta(-\Delta\tau)\rho_{in}m_I(\tau)m_I(\tau').
\eea
Then
\bea
\delta\rho^{(a)}(\tau,\tau')&=&\left(
\begin{array}{cc}
	(1-p)e^{-i\omega\Delta\tau} & 0\\
	0 & p\,e^{i\omega\Delta\tau}
\end{array}\right),\nn\\
\delta\rho^{(b)}(\tau,\tau')&=&-\theta(\Delta\tau)
\left(
\begin{array}{cc}
	pe^{i\omega\Delta\tau} & 0\\
	0 & (1-p)\,e^{-i\omega\Delta\tau}
\end{array}\right),\nn\\
\delta\rho^{(c)}(\tau,\tau')&=&-\theta(-\Delta\tau)
\left(
\begin{array}{cc}
	pe^{i\omega\Delta\tau} & 0\\
	0 & (1-p)\,e^{-i\omega\Delta\tau}
\end{array}\right).
\eea
Therefore we can write 
\beq
\delta\rho(\tau,\tau')=\left((1-p)e^{-i\omega\Delta\tau}-p\,e^{i\omega\Delta\tau}\right)\sigma_3,
\eeq
where $\sigma_3$ is the $Z$ Pauli matrix.  Hence, we can write the final
state of the qubit after interacted with the vacuum as 
\beq
\rho_I(t)=\rho_{in}+\delta p(t)\sigma_3
\eeq
where we have defined the change in the population of the excited state as
\beq
\delta p(t)=g^2\int_{-\bar\tau}^td\tau\int_{-\bar\tau}^td\tau'\left((1-p)e^{-i\omega\Delta\tau}-p\,e^{i\omega\Delta\tau}\right)
G^+_{\alpha}(\tau,\tau').
\label{deltap}
\eeq
Therefore, after the interaction with the quantum vacuum, the density operator of the qubit is given by
\beq
\rho_I(t)=\left(
\begin{array}{cc}
	p+\delta p(t) & 0\\
	0 & 1-(p+\delta p(t))
\end{array}
\right).
\eeq
One can see that the final increase of the population of the excited state $\delta p=\delta p(\bar\tau)$, Eq. (\ref{deltap}), is caused
by the interaction with the vacuum and this would give us a thermal-like response,
as predicted by Unruh~\cite{unruh1976}. We also realize that the change in the probability of the excited state
depends also on the initial population of this state $p$, on the energy gap $\omega$
of the qubit system and also on the acceleration $\alpha$ of the qubit
when interacting with the vacuum.

\section{Regularization by Lorentzian profile}
\label{ap:regularization}

\begin{figure}[t]
	\centering
		\includegraphics[scale=0.45]{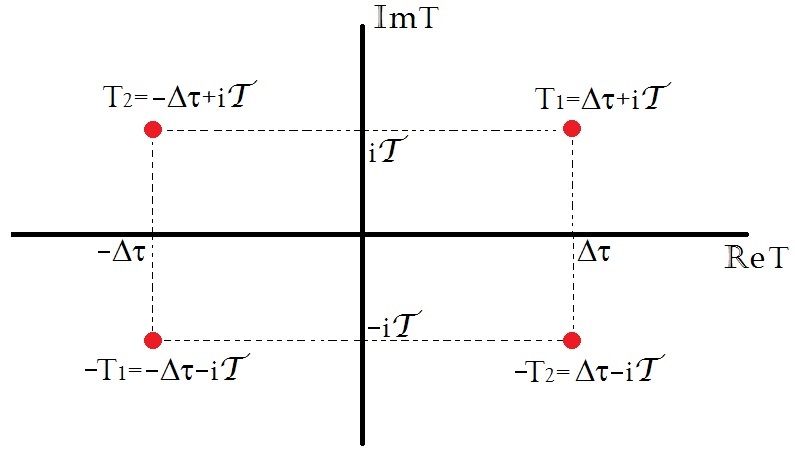}
		\caption{Pole structure of the regulator product  $\xi_{\cal T}(\tau)\xi_{\cal T}(\tau')$ in the complex plane of $T=\tau+\tau'$.}
	\label{4polos}
\end{figure}

\begin{figure}[ht]
	\centering
		\includegraphics[scale=0.35]{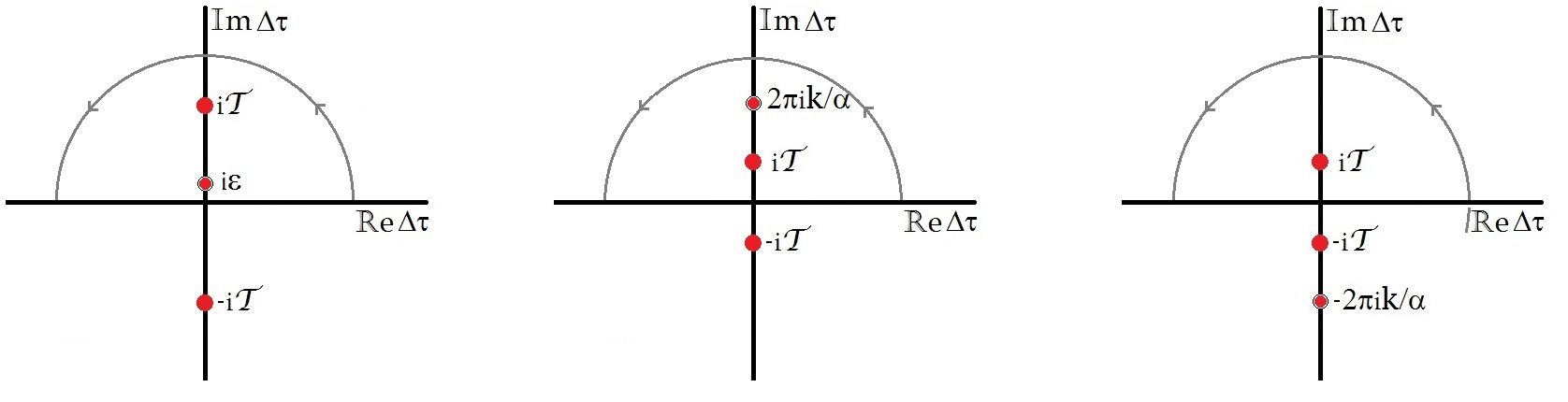}
		\caption{Pole structure in the complex plane of $\Delta\tau$ for the terms in Eq. (\ref{B11}).}
	\label{npolos}
\end{figure}

In this appendix, we will show more explicitly how the vacuum fluctuations transfer heat
to the qubit system. The integral that shows the connection between vacuum fluctuations and
the specific trajectory of the qubit is given by
\beq
{\cal J}(\alpha,\omega,{\cal T})=\int_{-{\cal T}/2}^{{\cal T}/2}d\tau\int_{-{\cal T}/2}^{{\cal T}/2}d\tau'e^{i\omega\Delta\tau}G^+_{\alpha}(\tau,\tau').
\label{calJ}
\eeq
Here, the finite-time interval ${\cal T}$ is defined in terms of the velocity $v$ and 
the acceleration $\alpha$. This finite-time integral can be extended to infinite by 
using a regulator that is compact in the domain we are interested in and vanish outside. 
This procedure is in the same spirit of the regularization by a cut-off function
of an infinite divergent integral. 
In particular, finite-time integrals related to the Unruh effect have been previously analyzed, e.g., in 
Refs.~\cite{Svaiter:92, Matsas:93}. 
However, since we are dealing with a kinematically closed trajectory
there is no need of taking the limit of infinite time.
Hence, we will identify Eq. (\ref{calJ}) with the expression
\beq
{\cal J}(\alpha,\omega,{\cal T})=\int_{-\infty}^{\infty}d\tau\int_{-\infty}^{\infty}d\tau'
\xi_{\cal T}(\tau)\xi_{\cal T}(\tau')e^{i\omega\Delta\tau}G^+_{\alpha}(\tau,\tau'),
\eeq
where $\xi_{\cal T}(\tau)$ is a compact smooth function that is nonvanishing 
for $-{\cal T}/2<\tau<{\cal T}/2$, being approximately zero outside that domain. 
We could try to adopt a Gaussian regulator of the kind $\xi_{gauss}\sim \exp(-\tau^2/{\cal T}^2)$.
However, as we expand this Gaussian regulator to the complex plane $\tau\rightarrow z\in\mathbb{C}$, 
we see that it does not vanish for any $|z|\rightarrow\infty$. Actually, depending on the phase in $z=|z|e^{i\theta}$, 
this regulator can be divergent.
In order to bypass this difficulty, we adopt as a regulator $\xi_{\cal T}(\tau)$ a Lorentzian profile 
rather than a Gaussian profile, i.e.
\beq
\xi_{\cal T}(\tau)=\frac{({\cal T}/2)^2}{\tau^2+({\cal T}/2)^2}.
\eeq
The expression above for $\xi_{\cal T}(\tau)$ yields
\beq
\xi_{\cal T}(\tau)\xi_{\cal T}(\tau')=\frac{{\cal T}^4}{(T^2-T_1^2)(T^2-T_2^2)},
\label{xixi}
\eeq
where we have defined $T_1=\Delta\tau+i{\cal T}$ and $T_2=-\Delta\tau+i{\cal T}$. 
Hence, Eq. (\ref{xixi}) shows $\xi_{\cal T}(\tau)\xi_{\cal T}(\tau')$ explicitly as a 
function of $T=\tau+\tau'$. This function posses a pole structure in the complex 
plane of $T$, as shown in Fig. (\ref{4polos}). Naturally, the polar structure depends 
on $\Delta\tau=\tau-\tau'$. 
Now, performing the change of integration
variables from $(\tau,\tau')$ to $(T,\Delta\tau)$ one gets
\beq
{\cal J}(\alpha,\omega,{\cal T})=\frac{1}{2}\int_{-\infty}^{\infty}dT\int_{-\infty}^{\infty}d(\Delta\tau)\,
\xi_{\cal T}(\tau)\xi_{\cal T}(\tau')e^{i\omega\Delta\tau}G^+_{\alpha}(\Delta\tau).
\label{calJ2}
\eeq
Here, we have made explicit the dependence on the Wightman correlation functions 
of the scalar field evaluated at an accelerated trajectory, which is given by
\beq
G^+_{\alpha}(\tau,\tau')=G^+_{\alpha}(\Delta\tau)=-\frac{\alpha^2}{16\pi^2}\frac{1}{\sinh^2(\frac{\alpha\Delta\tau}{2}-i\epsilon\alpha)}.
\eeq
Therefore, one can see that the only dependence on $T$ is given by the regulator factors 
$\xi_{\cal T}(\tau)\xi_{\cal T}(\tau')$, such that
\beq
{\cal J}(\alpha,\omega,{\cal T})=\frac{1}{2}\int_{-\infty}^{\infty}d(\Delta\tau)
e^{i\omega\Delta\tau}G^+_{\alpha}(\Delta\tau)\int_{-\infty}^{\infty}dT\,\xi_{\cal T}(\tau)\xi_{\cal T}(\tau').
\label{B7}
\eeq
Then, performing the integral on $T$ first, using the pole structure in Fig. (\ref{4polos}) and the Cauchy-Riemann
residue theorem of complex variables~\cite{Fisher:99} one gets
\beq
\int_{-\infty}^{\infty}dT\,\xi_{\cal T}(\tau)\xi_{\cal T}(\tau')=\frac{\pi}{2}\frac{{\cal T}^3}{\Delta\tau^2+{\cal T}^2}.
\eeq
Putting this result into Eq. (\ref{B7}) and using the alternative representation 
of the Wightman function on the accelerated trajectory~\cite{birrell}
\beq
G^+_{\alpha}(\Delta\tau)=-\frac{1}{4\pi^2}\sum_{k=-\infty}^{\infty}{\frac{1}{(\Delta\tau-i\epsilon-2\pi ik/\alpha)^2}},
\eeq
we get that
\beq
{\cal J}(\alpha,\omega,{\cal T})=-\frac{{\cal T}^3}{16\pi}\int_{-\infty}^{\infty}d(\Delta\tau)\,
\sum_{k=-\infty}^{\infty}\frac{e^{i\omega\Delta\tau}}{(\Delta\tau-i\epsilon-2\pi ik/\alpha)^2(\Delta\tau^2+{\cal T}^2)}.
\eeq
Splitting the term in the above sum, it follows that
\bea
{\cal J}(\alpha,\omega,{\cal T})&=&-\frac{{\cal T}^3}{16\pi}\int_{-\infty}^{\infty}d(\Delta\tau)\,
\bigg(
\frac{e^{i\omega\Delta\tau}}{(\Delta\tau-i\epsilon)^2(\Delta\tau^2+{\cal T}^2)}\nn\\
&&+\sum_{k=0}^{\infty}\frac{e^{i\omega\Delta\tau}}{(\Delta\tau-2\pi ik/\alpha)^2(\Delta\tau^2+{\cal T}^2)}\nn\\
&&+\sum_{k=0}^{\infty}\frac{e^{i\omega\Delta\tau}}{(\Delta\tau+2\pi ik/\alpha)^2(\Delta\tau^2+{\cal T}^2)}
\bigg).
\label{B11}
\eea
All the terms in the above expression have simple poles in $\pm i{\cal T}$. The first term corresponds to $k=0$
and possess a pole of second order at $i\epsilon$. In the second and third terms, we have neglected the $i\epsilon$
contribution, since in these terms the double pole is finite at $\pm 2\pi i/\alpha$, respectively.
This pole structure is shown in Fig. (\ref{npolos}).

By using the Residue theorem and the definition of the transcendental Lerch-Hurwitz function
\beq
\phi(z,s,a)=\sum_{k=0}^{\infty}\frac{z^k}{(k+a)^s},
\eeq
we obtain then that
\bea
\!\!\!\!\!\!\!\!\!\!\!\!\!\!{\cal J}(\alpha,\omega,{\cal T})&=&\frac{(\alpha{\cal T}/2)^2e^{-|\omega|{\cal T}}}{8\sin^2(\alpha{\cal T}/2-i\epsilon)}
+\frac{|\omega|{\cal T}}{4}\theta(\omega)\nn\\
&\!\!\!\!\!\!\!\!\!\!\!\!\!\!\!\!\!\!\!\!\!\!\!\!\!\!\!\!\!\!\!\!\!\!\!\!\!\!\!\!\!\!\!\!\!\!\!+&\!\!\!\!\!\!\!\!\!\!\!\!\!\!\!\!\!\!\!\!\!\!\!\!\!\!\frac{\alpha^2{\cal T}^2e^{-2\pi|\omega|/|\alpha|}}{32\pi^2}
\left(\phi(e^{-2\pi|\omega|/|\alpha|},2,1+\frac{|\alpha|{\cal T}}{2\pi})-\phi(e^{-2\pi|\omega|/|\alpha|},2,1-\frac{|\alpha|{\cal T}}{2\pi})\right)
\nn\\
&\!\!\!\!\!\!\!\!\!\!\!\!\!\!\!\!\!\!\!\!\!\!\!\!\!\!\!\!\!\!\!\!\!\!\!\!\!\!\!\!\!\!\!\!\!\!\!\!\!\!+&\!\!\!\!\!\!\!\!\!\!\!\!\!\!\!\!\!\!\!\!\!\!\!\!\!\!\frac{|\omega\alpha|{\cal T}^2e^{-2\pi|\omega|/|\alpha|}}{16\pi}
\left(\phi(e^{-2\pi|\omega|/|\alpha|},1,1+\frac{|\alpha|{\cal T}}{2\pi})-\phi(e^{-2\pi|\omega|/|\alpha|},1,1-\frac{|\alpha|{\cal T}}{2\pi})\right),
\label{calJ2}
\eea
the small imaginary part $i\epsilon$ in the first term of the equation above can be iterpreted using
the Cauchy principal value $1/(x\mp i\epsilon)=(P/x)\pm i\pi\delta(x)$, see \cite{birrell}. In this way, with $\epsilon>0$, we have that the function 
${\cal J}$ tends to zero in the limit of vanishing interaction time ${\cal T}$.
This function, Eq. (\ref{calJ2}), apparently depends on three independent variables. However, as one can check, 
the combination between these variables
leads us to a two-variable function. Indeed, if we define the function
\bea
J(x,y)&=&
\frac{(y/2)^2e^{-|x|y}}{8\sin^2(y/2)}-\frac{1}{8}+\frac{|x|y}{4}\theta(x)\nn\\
&+&
\frac{y^2e^{-2\pi|x|}}{32\pi^2}
\left(\phi(e^{-2\pi|x|},2,1+\frac{y}{2\pi})-\phi(e^{-2\pi|x|},2,1-\frac{y}{2\pi})\right)
\nn\\
&+&
\frac{|x|y^2e^{-2\pi|x|}}{16\pi}
\left(\phi(e^{-2\pi|x|},1,1+\frac{y}{2\pi})-\phi(e^{-2\pi|x|},1,1-\frac{y}{2\pi})\right),
\label{j}
\eea
then we can rewrite
\beq
{\cal J}(\alpha,\omega,{\cal T})=J\left(\frac{\omega}{\alpha},\alpha{\cal T}\right).
\eeq
By using this analytical expression, we can derive the general result shown in the Appendix~A.
There, we obtained that the final general correction to the population of the excited state of the 
qubit due to the vacuum fluctuations is
\beq
\delta p=g^2\int_{-\bar\tau}^{\bar\tau}d\tau\int_{-\bar\tau}^{\bar\tau} d\tau'\left((1-p)e^{-i\omega\Delta\tau}-p\,e^{i\omega\Delta\tau}\right)
G^+_{\alpha}(\tau,\tau').
\eeq
Then, using the definition of the integral in Eq. (\ref{calJ2}), we can write
\bea
\delta p&=&g^2\left((1-p){\cal J}(\alpha,-\omega,2\bar\tau)-p\,{\cal J}(\alpha,\omega,2\bar\tau)\right),\nn\\
&=&g^2\left((1-p)J\left(-\frac{\omega}{\alpha},2\alpha\bar\tau\right)-p\,J\left(\frac{\omega}{\alpha},2\alpha\bar\tau\right)\right).
\eea
Defining the reduced acceleration of the qubit as the ratio $a=\alpha/\omega$ and 
taking into account that the interaction time of the qubit with the vacuum have to be such $\bar\tau=\arctanh(v)/\alpha$,
we obtain
\bea
\delta  p/g^2&=&(1-p) J\left(-\frac{1}{a},2\arctanh{v}\right)-
p J\left(\frac{1}{a},2\arctanh{v}\right),\nn\\
&=& (1-2p)J\left(-\frac{1}{a},2\arctanh{v}\right)-
p\frac{\arctanh{v}}{2a}.
\label{B18}
\eea
In the final line we have used the property $J(x,y)-J(-x,y)=xy/4$, which follows from Eq. (\ref{j}). This 
concludes the analysis of the regularized perturbative expression for the excitation correction $\delta p$. 

\end{appendices}


\begin{thebibliography}{99}

\bibitem{Gemmer:book}  J. Gemmer, M. Michel, and G. Mahler, 
{\it Quantum Thermodynamics: Emergence of Thermodynamic Behavior Within Composite Quantum Systems}, 
Lecture Notes in Physics, Vol. {\bf 657}, Springer-Verlag Berlin Heidelberg, 2004. 

\bibitem{Goold16} J. Goold, M. Huber, A. Riera, L. del Rio, and P. Skrzypczyk, 
J. Phys. A: Math. Theor. {\bf 49}, 143001 (2016).

\bibitem{Maruyama:09} K. Maruyama, F. Nori, and V. Vedral, Rev. Mod. Phys. {\bf 81}, 1 (2009).

\bibitem{Scovil59} H. E. D. Scovil and E. O. Schulz-DuBois, Phys. Rev. Lett. {\bf 2}, 262 (1959).

\bibitem{Kieu04} T. D. Kieu, Phys. Rev. Lett. {\bf 93}, 140403 (2004).

\bibitem{Kieu06} T. D. Kieu, Eur. Phys. J. D {\bf 39}, 115 (2006).

\bibitem{Zhang:07} T. Zhang, W.-T. Liu, P.-X. Chen, and C.-Z. Li, Phys. Rev. A {\bf 75}, 062102 (2007). 

\bibitem{Zhang:08} G.-F. Zhang, Eur. Phys. J. D {\bf 49}, 123 (2008). 

\bibitem{Wang:09} H. Wang, S. Liu, and J. He, Phys. Rev. E {\bf 79}, 041113 (2009).  

\bibitem{bardeen1973} J. M. Bardeen, B. Carter and S. W. Hawking,
Commun. Math. Phys. {\bf 31}, 161 (1973).

\bibitem{bekenstein1973} J. D. Bekenstein, 
Phys. Rev. D {\bf 7}, 2333 (1973).

\bibitem{hawking1974} S. W. Hawking, 
Nature {\bf 248}, 30 (1974).

\bibitem{hawking1975} S. W. Hawking, Commun. Math. Phys. {\bf 43}, 199 (1975).

\bibitem{hawking1976} S. W. Hawking, Phys. Rev. D {\bf 13}, 191 (1976).

\bibitem{birrell} N. D. Birrell and P. C. W. Davis, \emph{Quantum Fields in Curved Space}, Cambridge University Press, New York, (1982).

\bibitem{unruh1976} W. G. Unruh, Phys. Rev. D {\bf 14}, 870 (1976).

\bibitem{Felicetti:15} S. Felicetti, C. Sab\'{\i}n, I. Fuentes, L. Lamata, G. Romero, and E. Solano, 
Phys. Rev. B {\bf 92}, 064501 (2015).

\bibitem{Laguna:17} J. Rodr\'{\i}guez-Laguna, L. Tarruell, M. Lewenstein, and A. Celi, Phys. Rev. A {\bf 95}, 013627 (2017).

\bibitem{Wang:14} J. Wang, Z. Tian, J. Jing, and H. Fan, Sci. Rep. {\bf 4}, 7195 (2014).

\bibitem{Ahluwalia:16} D. V. Ahluwalia, L. Labunb, and G. Torrieric, J. Phys.: Conference Series {\bf 706}, 042006 (2016).

\bibitem{Cozzella:17} G. Cozzella, A. G. S. Landulfo, G. E. A. Matsas, and D. A. T. Vanzella, 
Phys. Rev. Lett. {\bf 118}, 161102 (2017).

\bibitem{Crispino:08} L. C. B. Crispino, A. Higuchi, and G. E. A. Matsas, 
Rev. Mod. Phys. {\bf 80}, 787 (2008).

\bibitem{dewitt1979} B. S. DeWitt, {\em ``General relativity, an Einstein centenary survey ''} 
(Cambridge University Press, Great Britain, 1979), edited by S. W. Hawking and W. Israel.

\bibitem{Campisi:11} M. Campisi, P. H\"anggi, and P. Talkner, 
Rev. Mod. Phys. {\bf 83}, 771 (2011).

\bibitem{Horn:13} R. A. Horn and C. R. Johnson, {\it{Matrix Analysis}}, 2nd Edition, 
Cambridge University Press, New York, NY, USA, 2013. 

\bibitem{Born:28}
{M. Born and V. Fock}, 
{Z. Phys. {\bf 51}, 165 (1928)}.

\bibitem{Kato:50}
{T. Kato}, 
{J. Phys. Soc. Jpn. {\bf 5}, 435 (1950)}.

\bibitem{Messiah:book}
{A. Messiah}, 
\textit{Quantum mechanics}, 
{North-Holland}, {Amsterdam} (1962).

\bibitem{Niedenzu17} W. Niedenzu, V. Mukherjee, A. Ghosh, A. G. Kofman, G. Kurizki, 
e-print arXiv:1703.02911 (2017).

\bibitem{Fisher:99} S. D. Fisher, {\em ``Complex Variables''},  2nd Edition, Dover, USA, 1999.

\bibitem{Deffner:11} S. Deffner and E. Lutz, Phys. Rev. Lett. {\bf 107}, 140404 (2011). 

\bibitem{Svaiter:92} B. F. Svaiter and N. F. Svaiter, Phys. Rev. D {\bf 46}, 5267 (1992).

\bibitem{Matsas:93} A. Higuchi, G. E. A. Matsas, and C. B. Peres, Phys. Rev. D {\bf 48}, 3731 (1993).

\end{thebibliography}
\end{document}